\documentclass[10pt,conference]{IEEEtran}
\usepackage{cite}
\usepackage{amsmath,amssymb,amsfonts}
\usepackage{algorithmic}
\usepackage{graphicx}
\usepackage{textcomp}
\usepackage[hyphens]{url}
\usepackage{fancyhdr}
\usepackage{hyperref}
\usepackage{color,soul}
\usepackage{balance}

\newcommand{\hpcayear}{2025}
\usepackage[noabbrev,capitalize]{cleveref}
\usepackage{comment}
\usepackage{xspace}

\usepackage{tikz}
\usepackage{tcolorbox}
\usepackage{multirow}
\usepackage{enumitem}
\usepackage{tcolorbox}
\usepackage{booktabs}

\usepackage{xcolor}
\hypersetup{
    colorlinks,
    linkcolor={red!50!black},
    citecolor={blue},
    urlcolor={blue!80!black}
}

\newcommand*{\papername}{\textsc{Dapper}\@\xspace}
\newcommand*{\papernameS}{\textsc{Dapper-S}\@\xspace}
\newcommand*{\papernameH}{\textsc{Dapper-H}\@\xspace}

\DeclareRobustCommand\encircle[1]{\tikz[baseline=(char.base)]{\node[shape=circle,fill,inner sep=0pt] (char) {\textcolor{white}{#1}}}}

\newcommand{\ignore}[1]{}

\newcommand{\PERFATTACK}{\textit{Perf-Attack}}
\newcommand{\PERFATTACKS}{\textit{Perf-Attacks}}
\newcommand{\NRH}{$\text{N}_{\text{RH}}$}
\newcommand{\NMIT}{$\text{N}_{\text{M}}$}
\newcommand{\NRG}{$\text{N}_{\text{RG}}$}
\newcommand{\NGCT}{$\text{N}_{\text{GC}}$}

\newcommand{\TREFW}{tREFW}
\newcommand{\TREFI}{tREFI}
\newcommand{\tRESET}{\text{t$_{\text{reset}}$}}
\newcommand{\tLEFT}{\text{t$_{\text{left}}$}}
\newcommand{\tRC}{tRC}
\newcommand{\tRFC}{tRFC}
\newcommand{\tRRDS}{tRRD\_S}
\newcommand{\ACTMAX}{\text{ACT$_{\text{MAX}}$}}
\newcommand{\ATITER}{\text{AT$_{\text{iter}}$}}
\newcommand{\ATTIME}{\text{AT$_{\text{time}}$}}
\newcommand{\MAPPINGATTACK}{\textit{Mapping-Capturing}}
\newcommand{\MAPPINGAGNOGATTACK}{\textit{Mapping-Agnostic}}
\newcommand{\RHTATTACK}{\textit{RH-Tracker-based}}
\newcommand{\DRFMSB}{\text{DRFM$_{\text{sb}}$}}

\newcommand{\RFMSB}{\text{RFM$_{\text{sb}}$}}
\newcommand{\RFMAB}{\text{RFM$_{\text{ab}}$}}

\newcommand{\REF}{\text{REF}}
\newcommand{\ACT}{\text{ACT}}
\newcommand{\PRE}{\text{PRE}}
\newcommand{\topic}[1]{\noindent{}\textbf{#1}: }
\newcommand{\hpcasubmissionnumber}{255}
\title{DAPPER: A Performance-Attack-Resilient Tracker for RowHammer Defense\vspace{-0.1in}}
\def\hpcacameraready{} % Uncomment to build camera-ready version

\newcommand\hpcaauthors{Jeonghyun Woo and Prashant J. Nair}
\newcommand\hpcaaffiliation{Department of Electrical and Computer Engineering \\ The University of British Columbia}
\newcommand\hpcaemail{\{jhwoo36, prashantnair\}@ece.ubc.ca\vspace{-0.1in}}

\author{
  \ifdefined\hpcacameraready
    \IEEEauthorblockN{\hpcaauthors{}}
      \IEEEauthorblockA{
        \hpcaaffiliation{} \\
        \hpcaemail{}
      }
  \else
    \IEEEauthorblockN{\normalsize{HPCA \hpcayear{} Submission
      \textbf{\#\hpcasubmissionnumber{}}} \\
      \IEEEauthorblockA{
        Confidential Draft \\
        Do NOT Distribute!!
      }
    }
  \fi 
}

\fancypagestyle{camerareadyfirstpage}{%
  \fancyhead{}
  
  \fancyhead[C]{
    \ifdefined\aeopen
    \parbox[][12mm][t]{13.5cm}{\hpcayear{} IEEE International Symposium on High-Performance Computer Architecture (HPCA)}    
    \else
      \ifdefined\aereviewed
      \parbox[][12mm][t]{13.5cm}{\hpcayear{} IEEE International Symposium on High-Performance Computer Architecture (HPCA)}
      \else
      \ifdefined\aereproduced
      \parbox[][12mm][t]{13.5cm}{\hpcayear{} IEEE International Symposium on High-Performance Computer Architecture (HPCA)}
      \else
      \parbox[][0mm][t]{13.5cm}{\hpcayear{} IEEE International Symposium on High-Performance Computer Architecture (HPCA)}
    \fi 
    \fi 
    \fi 
    \ifdefined\aeopen 
      \includegraphics[width=12mm,height=12mm]{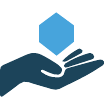}
    \fi 
    \ifdefined\aereviewed
      \includegraphics[width=12mm,height=12mm]{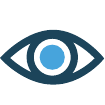}
    \fi 
    \ifdefined\aereproduced
      \includegraphics[width=12mm,height=12mm]{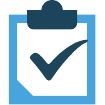}
    \fi
  }
  \fancyfoot[C]{}
}
\begin{document}
\maketitle

%Enables the camera ready header and footer
\ifdefined\hpcacameraready 
  \thispagestyle{camerareadyfirstpage}
  \pagestyle{plain}
\else
  \thispagestyle{plain}
  \pagestyle{plain}
\fi

\newcommand{\hpcaheight}{0mm}
\ifdefined\eaopen
\renewcommand{\hpcaheight}{12mm}
\fi

%%%%%%%%%%%%%%%%%%%%%%%%%%%%%%%%%%%%%%%%
%%%%%%%% -- PAPER CONTENT STARTS -- %%%%%%%%%

%%%%%% -- PAPER CONTENT STARTS-- %%%%%%%%
\begin{abstract}
RowHammer vulnerabilities pose a significant threat to modern DRAM-based systems, where rapid activation of DRAM rows can induce bit-flips in neighboring rows. To mitigate this, state-of-the-art host-side RowHammer mitigations typically rely on shared counters or tracking structures. While these optimizations benefit benign applications, they are vulnerable to Performance Attacks (\PERFATTACKS{}), where adversaries exploit shared structures to reduce DRAM bandwidth for co-running benign applications by increasing DRAM accesses for RowHammer counters or triggering repetitive refreshes required for the early reset of structures, significantly degrading performance.

In this paper, we propose secure hashing mechanisms to thwart adversarial attempts to capture the mapping of shared structures. We propose \papername{}, a novel low-cost tracker resilient to \PERFATTACKS{} even at ultra-low RowHammer thresholds. We first present a secure hashing template in the form of \papernameS{}. We then develop \papernameH{}, an enhanced version of \papernameS{}, incorporating double-hashing, novel reset strategies, and mitigative refresh techniques. Our security analysis demonstrates the effectiveness of \papernameH{} against both RowHammer and \PERFATTACKS{}. Experiments with 57 workloads from SPEC2006, SPEC2017, TPC, Hadoop, MediaBench, and YCSB show that, even at an ultra-low RowHammer threshold of 500, \papernameH{} incurs only a 0.9\%  slowdown in the presence of \PERFATTACKS{} while using only 96KB of SRAM per 32GB of DRAM memory.
\end{abstract}

\section{Introduction}\label{introduction}
Modern Dynamic Random Access Memories (DRAM) face severe security issues due to technology scaling. DRAM cells store data as charge, and at smaller technology nodes, rapid accesses can induce data leakage from neighboring cells, resulting in bit-flips that adversaries can exploit to manipulate data\cite{seaborn2015exploiting, kim2014architectural, gruss2016rowhammer, half-double}. This vulnerability, known as RowHammer (RH)\cite{kim2014flipping}, occurs when rapidly accessed rows (aggressor rows) cause bit-flips in neighboring rows (victim rows). One approach to addressing RH involves integrating low-cost trackers into the host-side memory controller\cite{hydra, olgun2023abacus, start_hpca24, comet}. These trackers enable targeted mitigative refreshes to victim rows. However, while effective for both RH attacks and benign applications, these trackers remain vulnerable to Performance Attacks (\PERFATTACKS{}). This paper aims to develop a host-side low-cost RH tracker that is resilient against \PERFATTACKS{} even at ultra-low RH thresholds (\NRH{}).

RH mitigations typically aim to refresh victim rows before adversarial activations reach the RH threshold (\NRH{}), the minimum number of activations required to induce bit-flips. Over the last decade, \NRH{} has sharply declined from around 70K\cite{kim2014flipping} in 2014 to 4.5K\cite{kim2020revisiting} in 2020, increasing the number of vulnerable rows and escalating hardware tracking costs. Advanced trackers like Hydra\cite{hydra}, ABACUS\cite{olgun2023abacus}, and CoMeT\cite{comet} reduce these costs by grouping rows and tracking collective activation counts. Alternatively, START\cite{start_hpca24} leverages the Last-Level Cache (LLC) to store RH counts. These strategies optimize performance at ultra-low \NRH{} (e.g., \NRH{} $\leq$ 1K) and prevent RH under malicious attacks.

\begin{figure}[h!]
    \vspace{-0.1in}
    \centering
    \includegraphics[width=0.4\textwidth,height=\paperheight,keepaspectratio]{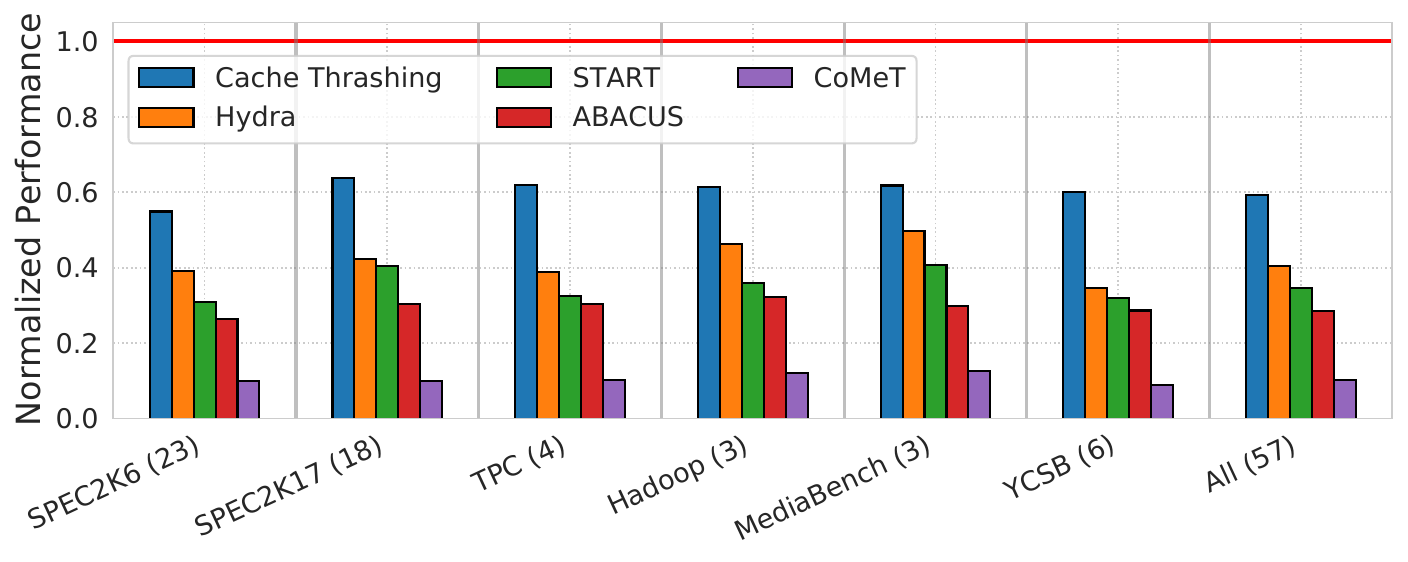}
    \vspace{-0.1in}
    \caption{Normalized performance of state-of-the-art host-side RowHammer (RH) mitigations\cite{hydra,comet,start_hpca24,olgun2023abacus} at the RH threshold of 500, evaluated under \RHTATTACK{} Performance Attacks (\PERFATTACK{}) and cache thrashing attacks on a dual-channel, dual-rank, DDR5 64GB system.  
    Our experiments with 57 workloads demonstrate that a single tailored \RHTATTACK{} \PERFATTACK{} application can reduce the performance of co-running applications by up to 90\%, while cache thrashing attacks degrade them by 40\% on average. This paper aims to develop a cost-effective \PERFATTACK{}-resistant RH tracker.}
    \label{fig:motivation}
    \vspace{-0.05in}
\end{figure}

However, grouping rows or using the LLC for tracking makes these mitigations vulnerable to \PERFATTACKS{}. Adversaries can force misses in RH counters or inflate group activation counts, leading to extra DRAM accesses and redundant refreshes that severely degrade the performance of co-running applications. These \PERFATTACKS{} exploit RH mitigation mechanisms and are more severe than attacks like cache thrashing. \cref{fig:motivation} shows that, at \NRH{} of 500, tailored \RHTATTACK{} \PERFATTACKS{} cause slowdowns of 60\% to 90\% across 57 workloads from SPEC2006\cite{SPEC2006}, SPEC2017\cite{SPEC2017}, TPC\cite{TPC}, Hadoop\cite{hadoop}, MediaBench\cite{MediaBench}, and YCSB\cite{ycsb}, whereas cache thrashing attacks result in only a 40\% slowdown. Ideally, we aim for a tracker that is resilient against \PERFATTACKS{}. This paper proposes \papername{}, a \PERFATTACK{}-resilient RH tracker that minimizes performance impact while providing secure defense against \PERFATTACKS{}.

To understand the design of \papername, it is crucial to decipher the hardware mechanisms exploited by \PERFATTACKS{}.
\begin{itemize}[leftmargin=*]
\item \textbf{Triggering Additional RH Counter Accesses}: Hydra\cite{hydra} uses shared group counters to track multiple rows with a single counter, reducing storage overhead. Hydra switches to per-row tracking for improved performance when a group counter reaches its thresholds. RH counters are stored in a reserved DRAM area, with a subset cached in the Row Counter Cache (RCC). When the RCC is full, and a miss occurs, Hydra evicts a stored counter and fetches a new one from DRAM. Similarly, START\cite{start_hpca24} stores individual RH counters directly in the LLC without grouping. Both Hydra's RCC and START's LLC can be overwhelmed by a tailored access pattern, leading to frequent evictions and counter fetches from DRAM. These repeated operations significantly degrade DRAM bandwidth and introduce substantial performance overhead for co-running applications.
\item \textbf{Triggering Redundant Refreshes}: CoMeT\cite{comet} uses a Count-Min Sketch-based tracker to share counters across rows, reducing storage overhead. However, due to counter sharing, it cannot reset counters after performing a mitigative refresh. CoMeT employs the Recent Aggressor Table (RAT) to track and selectively reset recently mitigated rows to address this. Performance degradation occurs when the number of aggressor rows exceeds the RAT's capacity, and those rows are frequently activated. Similarly, ABACUS\cite{olgun2023abacus} uses a Misra-Gries (MG) tracker shared across all banks, with its size designed to handle the maximum number of aggressor rows in a single bank during a refresh interval. However, the MG tracker’s spillover counter can overflow under adversarial patterns. CoMeT and ABACUS implement early preventive refreshes that mitigate this problem by refreshing all DRAM rows and resetting the structure, but an attacker can exploit these mechanisms to reduce DRAM bandwidth and cause slowdowns.
\end{itemize}
\vspace{0.01in}
\noindent To address these concerns, we introduce two versions of \papername, each designed to be built on top of the other.

\smallskip
\noindent\textbf{1. \papernameS{}}: We propose \papernameS as a simple mechanism to prevent RH counter access and mitigate \PERFATTACKS{}. \papernameS{} groups rows and assigns dedicated counters, called row group counters (RGCs), to each group. When an RGC reaches the Mitigation threshold (\NMIT{}), which is half of \NRH{}, it refreshes victim rows and resets the counter. Unlike previous approaches\cite{hydra, start_hpca24}, \papernameS stores all RGCs in an SRAM-based table within the memory controller, avoiding main memory storage. Rows are randomly assigned to groups using a secure hash with periodically updated keys, reducing DRAM accesses and preventing attackers from targeting specific rows. This randomization minimizes the risk of overwhelming an RGC. Hash keys are refreshed every 12$\mu$s to effectively mitigate \PERFATTACKS{} with some overhead. However, \papernameS{} remains vulnerable to two \MAPPINGAGNOGATTACK{} attacks: (1) a `streaming' attack, where all rows are activated, overwhelming all RGCs and triggering redundant refreshes, and (2) a `refresh attack' that repeatedly activates one or more rows, causing mitigative refreshes for all rows sharing the same RGC. Despite this, \papernameS is a robust foundation for further enhancements to strengthen the tracker.

\smallskip
\noindent\textbf{2. \papernameH}: We develop \papernameH{} to address the limitations of \papernameS. \papernameH uses two SRAM tables for storing RGCs, each with a separate secure hash. The mitigation process occurs only when both RGCs reach the \NMIT{}, which is also half of \NRH{}, similar to \papernameS{}. This means that an attacker must identify a row's RGCs in both tables to overwhelm the counts. Additionally, \papernameH{} integrates a per-bank bit-vector for one of the RGC tables. This bit-vector acts as a filter, effectively managing scenarios where rows within the same group are activated across different banks. If the corresponding bit is unset, indicating that no row within this bank has been previously activated, \papernameH{} sets the bit without incrementing the RGC. Combined with the periodic rehashing of RGCs at each refresh window (\TREFW{}), the design effectively thwarts \MAPPINGATTACK{} attacks, which aim to capture even a single pair of row group mappings and exploit it, thereby avoiding performance degradation.

\noindent\textbf{Contributions}: This paper makes four key contributions
\begin{enumerate}[leftmargin=*]
    \item It demonstrates tailored \RHTATTACK{} \PERFATTACKS{} for state-of-the-art low-cost trackers. These attacks exploit shared structures like tracking counters or the LLC, which is used to minimize slowdowns in benign workloads.
    \item It identifies the causes of \PERFATTACKS{} in scalable RH trackers: additional accesses to in-DRAM RH counters and additional repetitive refreshes to DRAM for early structure resets. Proposes \papername to counter this issue.
    \item It introduces a naive solution, \papernameS, which randomizes rows mapped into shared counters for tracking using a single secure hash function. \papernameS makes it challenging for adversaries to learn the mapping and orchestrate \PERFATTACKS{}. Although vulnerable to mapping-agnostic streaming and refresh attacks, \papernameS provides a foundation to strengthen the tracker.
    \item It proposes \papernameH, an enhanced version of \papernameS, which uses double-hashing, a novel reset method, and a per-bank bit-vector. This enhancement makes it extremely challenging for adversaries to understand the mappings into the shared counters. Regardless of the access pattern, \papernameH is generally resilient against \PERFATTACK{} in 99.99\% of the refresh intervals and effectively mitigates mapping-agnostic \PERFATTACKS{}.
\end{enumerate}

Our experiments show that even at \NRH{} of 500, \papernameH incurs only a 0.9\% slowdown even under active \PERFATTACKS{} while using only 96KB of SRAM per 32GB of memory.

\section{Background}\label{background}
\subsection{Memory Organization and Timing Parameters}\label{dram_org}
A DRAM-based memory system is organized hierarchically. A Memory controller (MC) manages one or more channels, with each channel containing ranks composed of banks that operate in parallel. Each bank consists of a two-dimensional array of DRAM cells arranged in rows and columns. To access data, the MC issues an ACTIVATE (\ACT{}) command to load data into the row buffer. Switching to a different row within the same bank requires a PRECHARGE (\PRE{}) command to close the current row, followed by another \ACT{}. Due to charge leakage, DRAM performs periodic refreshes (\REF{}s) within a refresh window (\TREFW{}), typically 32ms in DDR5 systems\cite{micron_ddr5, jedec_ddr5_prac}. During \TREFW{}, the MC issues 8K auto-refresh commands at 3.9$\mu$s interval (\TREFI{}), refreshing multiple rows in each bank during refresh cycle time (\tRFC{})\cite{avatar}.

The row cycle time (\tRC{}) sets the minimum interval between \ACT{}s to different rows within the same bank. With \tRC{} of 48ns, a bank can undergo up to 616K activations within \TREFW{}. Similarly, tRRD defines the minimum time between \ACT{}s to different banks, with tRRD\_L applying to the same bank group and \tRRDS{} to different groups. For DDR5-6400 devices, the MC can issue up to 11.8M \emph{ACT}s per rank (32 banks) with \tRRDS{} latency. Finally, the write recovery time (tWR) defines the required wait time before issuing a \PRE{} after a write operation.

\subsection{The RowHammer Vulnerability}\label{bg:rowhammer}
RowHammer (RH) is a read-disturbance phenomenon where rapid activation of aggressor rows causes bit-flips in adjacent victim rows\cite{kim2014flipping}. The RH threshold (\NRH{}) is the minimum number of activations required to trigger bit-flips, which has declined significantly with DRAM technology scaling---from 70K in 2014\cite{kim2014flipping} to 4.5K by 2020\cite{kim2020revisiting}. Projections suggest \NRH{} could drop below 1K, with recent studies focusing on \NRH{} of 500 or lower\cite{hydra, comet, PrIDE, MINT}. While RH has been a critical security threat\cite{seaborn2015exploiting, gruss2018another, cojocar2019exploiting, kwong2020rambleed, half-double, jattke2022blacksmith, zenhammer_usenex_sec24, yao2020DeepHammer}, it also poses a reliability risk\cite{moesi-prime,archshield,xed} at low \NRH{}. 

\subsection{Threat Model}\label{threat_model}
We consider a DRAM-based system vulnerable to RH attacks. Attackers can execute malicious applications with \emph{user} privileges, such as Blacksmith\cite{jattke2022blacksmith}, to induce bit-flips in critical data structures such as page tables\cite{seaborn2015exploiting}. An attack succeeds \emph{any} DRAM row exceeds the RH threshold (\NRH{}) within \TREFW{}, causing bit-flips in adjacent rows\cite{safeguard}. 

Additionally, attackers can degrade the performance of co-running applications by launching Performance Attacks (\PERFATTACKS{}). In \PERFATTACKS{}, adversaries execute malicious workloads on one or more cores while benign workloads run on others. The RowPress\cite{luo2023rowpress} attack is out of scope as its effects are orthogonal. RowPress can be mitigated by limiting row open time\cite{luo2023rowpress} or incrementing activation counters based on the row open time\cite{Impress}.

\section{Motivation}\label{sec:motivation} 
\begin{figure*}[t!]
    \centering
    \includegraphics[width=1.75\columnwidth]{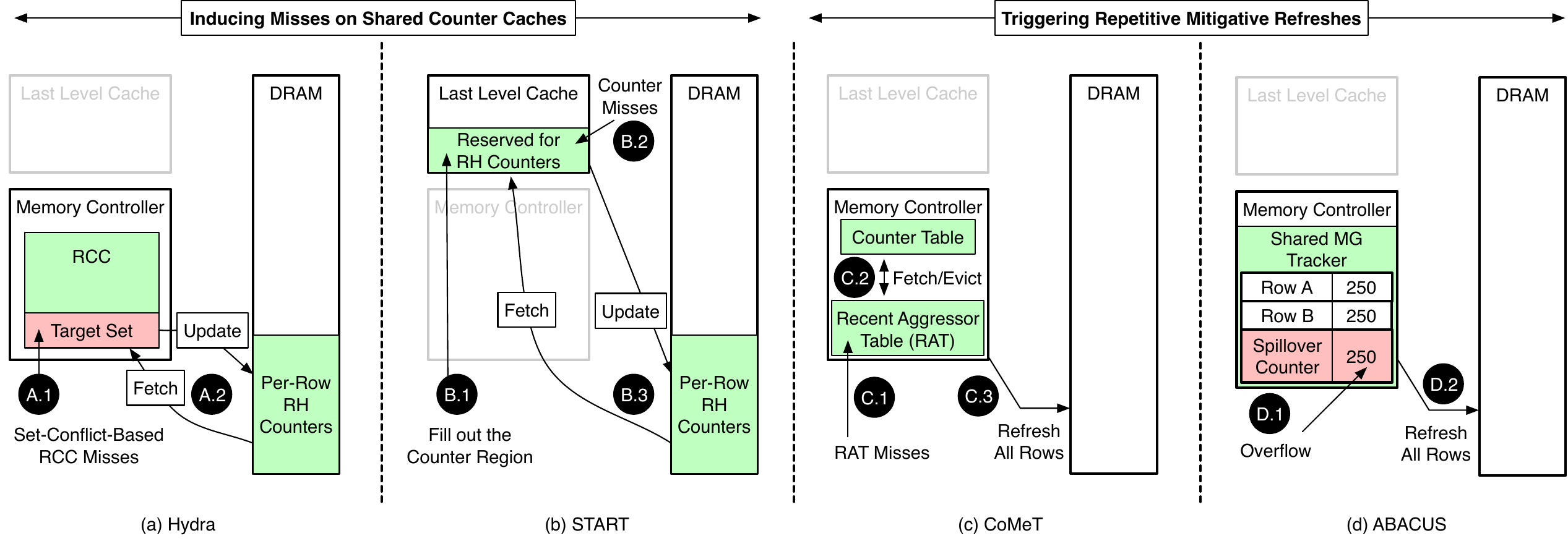}
    \vspace{-0.1in}
    \caption{An overview of the \RHTATTACK{} Performance Attacks (\PERFATTACK{}) tailored for state-of-the-art host-side low-cost RowHammer tracking mechanisms: Hydra\cite{hydra}, CoMeT\cite{comet}, START\cite{start_hpca24}, and ABACUS\cite{olgun2023abacus}. These attacks induce additional memory accesses or repetitive mitigative refreshes.}
    \label{fig:attack_explanation}
    \vspace{-0.05in}
\end{figure*}

\begin{figure*}[t!]
    \centering
    {\includegraphics[width=1.9\columnwidth]{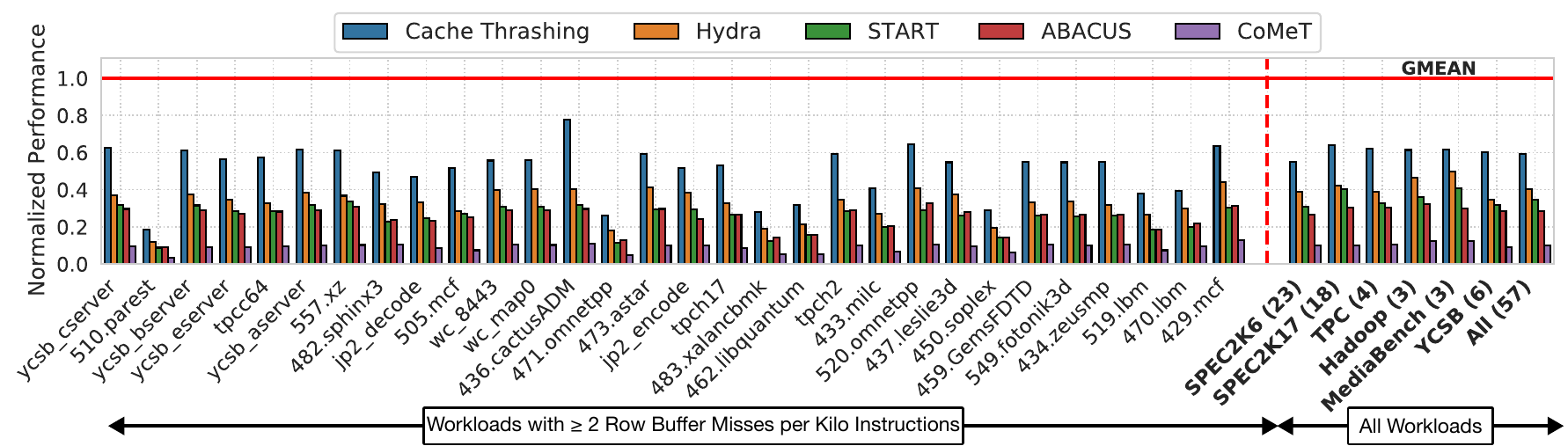}}
    \vspace{-0.1in}
    \caption{
    The performance impact of state-of-the-art RowHammer (RH) trakcers: Hydra\cite{hydra}, CoMeT\cite{comet}, START\cite{start_hpca24}, and ABACUS\cite{olgun2023abacus} under cache thrashing and tailored \RHTATTACK{} \PERFATTACKS{}. The performance of three benign applications is normalized to a baseline with no RH mitigation. On average, we notice a performance drop of 60\% to 90\% performance loss under \PERFATTACKS{} and a 40\% performance drop under the cache thrashing attack.
    }
    \label{fig:dos_results}
    \vspace{-0.13in}
\end{figure*}
State-of-the-art scalable RH mitigations use shared counters or tracking structures to reduce storage overhead. While these mitigations are effective on benign applications, their shared nature makes them vulnerable to Performance Attacks (\PERFATTACKS{}). This section describes each scalable mitigation, then introduces targeted \RHTATTACK{} \PERFATTACKS{}. We demonstrate their vulnerability even at high \NRH{} and conclude by showing that increasing cache size or memory channels alone is insufficient to prevent these attacks.

\subsection{Scalable RowHammer Mitigations}
\topic{Hydra\cite{hydra}} Hydra consists of three main components: the Group Counter Table (GCT), Row Counter Cache (RCC), and Row Counter Table (RCT). The GCT contains group counters (GCs), each shared among multiple rows, to track activations until they reach the predefined group counter threshold (\NGCT{}). This threshold is set to 80\% of the mitigation threshold (\NMIT{} $= \frac{\text{N}_{\text{RH}}}{2}$). Once a GC reaches \NGCT{}, Hydra transitions to per-row tracking for precise tracking. To minimize storage overhead in the memory controller, Hydra stores per-row RH counters in a reserved DRAM region (RCT) and caches a subset of these counters in the RCC, a small cache within the memory controller. When an RCC miss occurs, Hydra performs additional DRAM accesses to fetch and update RH counters, which can result in performance penalties. In our evaluation, we configure Hydra based on its original design\cite{hydra}, with a GC size of 128, 4K RCC entries per rank, and a 32-way set-associative cache with a random eviction policy.

\smallskip
\topic{START\cite{start_hpca24}} START dynamically allocates per-row RH counters within a reserved portion of the Last-Level Cache (LLC), eliminating the need for additional counter structures. The reserved LLC region, configured as half of the LLC, stores these counters. If this reserved region cannot accommodate all RH counters, START stores these counters in a reserved DRAM space, using the LLC region as a cache, similar to Hydra. This design can result in significant performance overhead due to two factors: 1) reducing the effective LLC capacity by dedicating part of the LLC to RH counter storage and 2) introducing additional DRAM accesses to fetch and update RH counters when counters are not cached in the LLC. Our evaluated system has 8 million RH counters, exceeding the 4 million counters that can fit in the reserved LLC region. Thus, we configure START to store RH counters in DRAM and use the reserved LLC region as a cache for these counters.
\
smallskip
\topic{CoMeT\cite{comet}} CoMeT leverages a Count-Min Sketch technique for its counter table (CT), enabling efficient counter sharing across multiple rows and reducing storage overhead. However, counter-sharing prevents individual counters from being reset until the next periodic reset. To address this, CoMeT integrates a Recent Aggressor Table (RAT) to track rows recently involved in mitigations. The RAT uses per-row counters to reduce unnecessary mitigations caused by saturated counters. Despite these optimizations, frequent RAT misses can lead to excessive refreshes as CoMeT resets its structures by refreshing all DRAM rows in the rank, resulting in significant slowdowns.
In our evaluation, we configure CoMeT following the original design\cite{comet}. The CT uses four hash functions, each managing 512 counters, while the Mitigation threshold is set to $\frac{\text{N}_{\text{RH}}}{4}$. The RAT is sized at 128 entries. CoMeT resets its structures every $\frac{\text{tREFW}}{3}$ by refreshing all DRAM rows. A 256-entry miss history is also employed, and extra resets are triggered when the RAT miss rate exceeds 25\%.

\smallskip
\topic{ABACUS\cite{olgun2023abacus}} ABACUS uses a single Misra-Gries (MG) tracker shared across all banks in the channel. To prevent counter overestimation and unnecessary mitigative refreshes, ABACUS introduces a per-bank bit-vector. The size of the MG tracker is determined by the maximum possible number of aggressor rows within a single bank during \TREFW{} at a given \NRH{}. For instance, at \NRH{} of 500, the MG tracker is configured with 2466 entries. However, since the MG tracker is shared across all banks, the counter can become saturated and remain unchanged until the periodic reset (every \TREFW{}). ABACUS addresses this by refreshing all DRAM rows in the channel and resetting the counters upon overflow, which introduces significant overhead. In our evaluation, we configure MG tracker sizes 309, 617, 1233, 2466, 4931, and 9783 entries for \NRH{} of 4K, 2K, 1K, 500, 250, and 125, respectively.

\subsection{Performance Attacks on Scalable RH Mitigations}\label{subsec:Perf-Attacks_on_SOTA}
To better understand the vulnerability of scalable RH mitigations to Performance Attacks (\PERFATTACKS{}), we develop tailored attack patterns that generate excessive DRAM accesses for updating RH counters or trigger numerous additional refreshes to reset tracking structures. \cref{fig:attack_explanation} provides an overview of these attacks. Unlike cache thrashing attacks, which degrade the performance of co-running applications by inducing frequent cache misses, \RHTATTACK{} \PERFATTACKS{} specifically exploit RH mitigation mechanisms. These attacks reduce the effective DRAM bandwidth for co-running benign applications by increasing access to in-DRAM RH counters—read and read-modify-write operations—or inducing frequent mitigative refreshes. As shown in \cref{fig:dos_results}, \RHTATTACK{} \PERFATTACKS{} can cause 16\% to 49\% more significant slowdowns compared to cache thrashing attacks.

\smallskip
\noindent\textbf{1. Inducing Misses on Shared Counter Caches}: \cref{fig:attack_explanation}(a) illustrates our proposed attack on Hydra, designed to trigger frequent counter updates by causing misses in the Row Counter Cache (RCC). By exploiting the RCC's set-associative nature, we target over 32 rows that map to the same RCC set. Activating 64 rows across different banks simultaneously leads to an 87\% miss probability in the RCC\footnote{With a 32-way RCC and random eviction, each set has a miss probability of $\frac{1}{32}$. Targeting $T$ rows results in a miss probability of $1 - (1 - \frac{1}{32})^T$.} (\encircle{A.1}). Each activation incurs two additional DRAM requests: one read to fetch the RH counter and one write to update the evicted counter (\encircle{A.2}). As shown in \cref{fig:dos_results}, this attack drops performance by 61\% on average, with a maximum of 88\% for \emph{510.parest}.

\cref{fig:attack_explanation}(b) depicts our attack against START, which reduces the LLC's effective capacity for benign applications. It also increases DRAM requests for RH counters, reducing DRAM bandwidth for co-executing benign workloads. By streaming access across all DRAM rows, the attack initially fills the LLC’s reserved RH counter regions (\encircle{B.1}). As the reserved LLC region becomes full, counter misses occur more frequently (\encircle{B.2}), leading to additional reads and writes for counter fetch and updates (\encircle{B.3}). On average, this attack drops performance by 65\%. START suffers significantly in memory-intensive workloads due to diminished LLC capacity and reduced DRAM bandwidth for benign applications caused by frequent RH counter updates. For example, \emph{510.parest} experiences a 91.2\% slowdown.

\smallskip
\noindent\textbf{2. Triggering Repetitive Mitigative Refreshes}: \cref{fig:attack_explanation}(c) illustrates our \PERFATTACK{} on CoMeT, targeting the limited 128-entry capacity of the Recent Aggressor Table (RAT). By rapidly activating more rows than the RAT can handle, such as 192 rows, the attack disrupts CoMeT's operations in two ways. First, it causes frequent evictions between the Counter Table (CT) and the RAT (\encircle{C.1}), leading to counter overestimation due to the CT's lack of a reset feature. This results in premature mitigative refreshes (\encircle{C.2}). Second, repeated RAT evictions force CoMeT to refresh all rows (\encircle{C.3}) to reset its structures, each lasting approximately 2.4 ms and imposing significant overhead. Our attack significantly increases the number of resets, leading to an average performance loss of 90\%. CoMeT is particularly vulnerable to \PERFATTACK{}, as the reset operation can occur every 1 ms, blocking access for 2.4 ms each time.

\cref{fig:attack_explanation}(d) shows our \PERFATTACK{} on ABACUS, which targets the shared N-entry Misra-Gries tracker. By sequentially activating rows across different banks (e.g., row 0 in bank 0, row 1 in bank 1), we overflow the spillover counter every $N \times \frac{T_{RH}}{2}$ activations (\encircle{D.1}). For example, with \NRH{} of 500 and 2466 tracker entries, this strategy can force a reset and induce refreshes for all DRAM rows within 2 ms (\encircle{D.2}). This Attack reduces performance by an average of 72\%, with up to a 91\% slowdown.

\begin{figure}[b!]
    \vspace{-0.2in}
    \centering
    \includegraphics[width=0.4\textwidth,height=\paperheight,keepaspectratio]{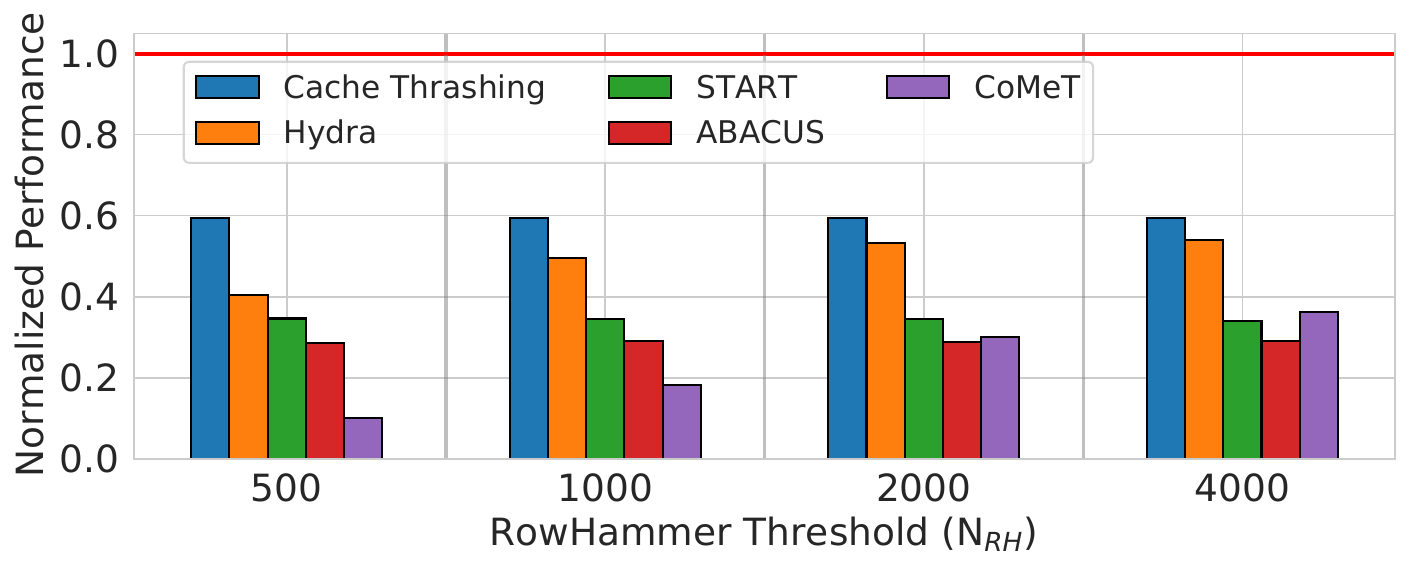}
    \vspace{-0.1in}
    \caption{Normalized performance of scalable RowHammer (RH) mitigations under cache thrashing and \RHTATTACK{} \PERFATTACKS{} as the RH threshold (\NRH{}) varies. Even at \NRH{} of 4K, scalable mitigations exhibit significant slowdowns of 46\% to 71\% under \RHTATTACK{} \PERFATTACKS{}, which is 5\% to 30\% higher than the slowdowns caused by cache thrashing attacks.
    }
    \label{fig:DoS-trh-sensitivity}
\end{figure}
\subsection{Attack Sensitivity to RowHammer Thresholds}
\cref{fig:DoS-trh-sensitivity} shows the normalized performance of scalable RH mitigations under cache thrashing and \RHTATTACK{} \PERFATTACKS{} as \NRH{} varies. Even at \NRH{} of 4K, scalable mitigations experience significant slowdowns from 46\% to 71\%, compared to a 41\% slowdown caused by cache thrashing attacks. Hydra and CoMeT experience higher slowdowns as \NRH{} decreases because fewer activations are needed to cause additional DRAM accesses or trigger early resets with repetitive refreshes. In contrast, START and ABACUS show consistent slowdowns across evaluated \NRH{}. This is because \PERFATTACKS{} on START and ABACUS involve streaming activations across multiple rows, consistently triggering additional DRAM accesses for counter updates and early reset with repetitive refreshes, which are inherently independent of \NRH{}. These findings underscore the persistent vulnerability of scalable RH mitigations to \PERFATTACKS{}, even at high \NRH{}.

\subsection{Attack Sensitivity to LLC Capacity and Memory Channels}
\begin{figure}[h]
    \vspace{-0.15in}
    \centering    \includegraphics[width=0.4\textwidth,height=\paperheight,keepaspectratio]{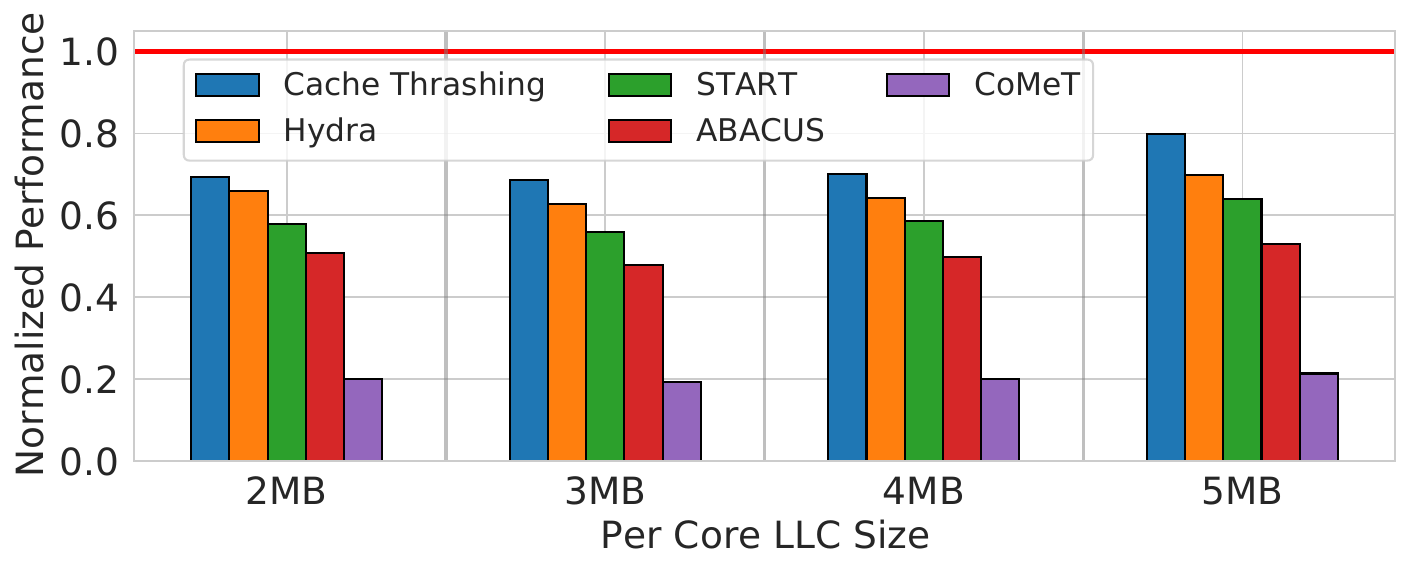}
    \vspace{-0.13in}
    \caption{
    Normalized performance of scalable RowHammer (RH) mitigations under cache thrashing and \RHTATTACK{} \PERFATTACKS{} with eight memory channels and the RH threshold (\NRH{}) of 500, as the per-core LLC size varies. Even with a 5MB per-core LLC, scalable mitigations exhibit slowdowns of 30\% to 79\%, which is 10\% to 59\% more severe than cache thrashing attacks.
    }
    \label{fig:DoS-LLC-sensitivity}
    \vspace{-0.18in}
\end{figure}

We use a four-core system with a 2MB per-core Last-Level Cache (LLC) and two memory channels, each with 32GB DDR5 memory per channel (64GB total), as our baseline configuration. Modern systems often feature larger memory capacities, additional channels, and larger LLC sizes per core. For instance, recent Intel Xeon processors support up to 12 memory channels with four ranks per channel, while AMD EPYC processors include LLC sizes of 2MB or more per core. To evaluate the \PERFATTACKS{} on these system configurations, we simulate a system with eight memory channels with 64GB per-channel memory (512GB total), and vary the per-core LLC size from 2MB to 5MB (8MB to 20MB total). \cref{fig:DoS-LLC-sensitivity} shows that \PERFATTACKS{} drops performance by 30\% to 79\%, compared to a 20\% slwdown caused by cache thrashing attacks. These results highlight that, even with large LLC capacities and increased memory channels, current scalable RH mitigations remain highly susceptible to \PERFATTACKS{}.

\subsection{Attack Potency Without Internal Knowledge}
We have assumed that an attacker knows internal details, such as the structure sizes of each RH mitigation. However, the attacker can maintain attack potency even without this information. The proposed \PERFATTACKS{} for START and ABACUS do not require internal details, as they only need to activate different rows with each attempt. Although the attack for Hydra requires knowledge of the Row Counter Cache (RCC) mapping to cause set conflicts and incur redundant DRAM accesses for counter updates, randomly selecting N rows and repeatedly activating them can achieve similar attack potency. Due to the limited capacity of the RCC, this attack will eventually fill the RCC, resulting in capacity misses instead of set-conflict misses. We validate that this attack causes slowdowns comparable to the originally proposed attack. For CoMeT, the attacker can randomly select N rows, repeatedly activate them, and monitor for early resets, which are easily detectable as they block DRAM access for 2.4 ms. Repeating this process several times, the attacker can deduce the Recent Aggressor Table (RAT) sizes and the number of aggressor rows needed to trigger \PERFATTACKS{}. This step is required only once, after which the attacker can continuously launch \PERFATTACKS{}. Thus, scalable RH mitigations remain highly susceptible, even in environments with limited attack knowledge.

\section{Evaluation Methodology}\label{sec:eval_method}
\smallskip
\topic{Simulation Framework}We evaluate designs using the publicly available cycle-accurate memory simulator Ramulator\cite{kim2015ramulator,ramulator_opensource, ramulator2}. Our simulated system configuration is detailed in \cref{table:system_config}. Our baseline setup consists of four out-of-order cores with a shared Last-Level cache (LLC). We use an out-of-order core model implemented in Ramulator similar to prior research\cite{olgun2023abacus,yauglikcci2022hira, qprac, breakhammer2024}. The LLC is a 16-way set-associative cache with an 8MB capacity. Our baseline system has two memory channels, each with a 32GB DDR5 DIMM. We simulate DDR5 6400MT/s DRAM\cite{micron_ddr5}, which has four banks per bank group and eight bank groups per rank. Each bank contains 64K rows that are 8KB in size. Within a 32ms refresh window (\TREFW{}), a single bank can experience up to approximately 616K activations, while a memory channel can encounter up to 11.8M activations. 

\begin{table}[h!]
\vspace{-0.1in}
\begin{center}
\begin{small}
\caption{System Configuration}{
\vspace{-0.1in}
\resizebox{0.95\columnwidth}{!}{
\begin{tabular}{|c|c|}
\hline
  Processor           & 4 cores (OoO), 4GHz, 4-wide, 128 entry ROB\\\hline
  Last Level Cache (Shared)    & 8MB, 16-Way, 64B lines \\ \hline
  Memory size                  & 64 GB -- DDR5 \\
  Memory bus speed             & 3.2 GHz (6.4GHz DDR) \\
  tRCD-tRP-tCL & 16-16-16 ns\\
  \tRC{}, \tRFC{}, \TREFI{}      & 48ns, 295 ns, 3.9$\mu$s \\
  DRAM Organization     & 4 Banks x 8 Groups x 2 Ranks x 2 Channels \\
  Rows per bank, Size                & 64K, 8KB \\ \hline 
\end{tabular}}
\label{table:system_config}
}
\end{small}
\end{center}
\vspace{-0.1in}
\end{table}

\topic{Workloads and Configuration} We use 57 open-sourced applications\cite{ramulator_opensource} from SPEC2006\cite{SPEC2006}, SPEC2017\cite{SPEC2017}, TPC\cite{TPC}, Hadoop\cite{hadoop}, MediaBench\cite{MediaBench}, and YCSB\cite{ycsb} benchmark suites. We run four homogeneous workloads until each core completes 500 million instructions. For the RowHammer threshold (\NRH{}), we use a default value of 500 and conduct a sensitivity study with \NRH{} values from 125 to 4K. For mitigative actions (refreshes), we assume the memory controller uses commands such as Victim Row Refresh (VRR) to refresh victim rows adjacent to the aggressor row on a per-bank basis, following prior work\cite{hydra, olgun2023abacus, comet, kim2014flipping}. By default, each VRR command refreshes one victim row on each side of the aggressor row (i.e., blast radius (BR) of 1). Additionally, we evaluate our design with BR of 2 and the existing Same-Bank Directed Refresh Management (\DRFMSB{}) command.

\section{Designing the DAPPER Tracker}\label{sec:design}
\subsection{The Need for a Secure Hash-Based Mapping}
\papername employs a shared Row Group Counter (RGC) for low-cost tracking. For example, grouping 256 rows requires only 16KB of SRAM for a 32GB memory. However, relying solely on shared counters causes vulnerabilities to adversarial patterns (See \cref{subsec:Perf-Attacks_on_SOTA}). Attackers can uniformly activate rows to overestimate the RGC, triggering unnecessary mitigative refreshes even if individual rows have not reached the Mitigation threshold (\NMIT{}), which is half of \NRH{}. For example, with \NMIT{} of 250 and a group size of 256 rows (our default), activating each row in the group once will quickly reach \NMIT{} and falsely trigger the RH mitigations for RGC, leading to issue mitigative refreshes for all 256 rows in the group. 

\subsection{DAPPER-S: Using a Dynamic Secure Hash}
\papernameS{} mitigates the risk of overestimating Row Group Counters (RGCs) by employing a secure hash to randomize row group mappings. Specifically, it randomizes rows within a defined level, such as per rank (e.g., 2M rows in our baseline), ensuring an even distribution of counter updates. This randomization prevents rapid counter increments in specific RGCs, thus avoiding unnecessary mitigative refreshes, particularly in workloads with high spatial locality. For the remainder of this paper, we will use the per-rank (2M rows) level mapping as our default configuration.

The hash function for this randomization must be reversible to support mitigation operations. \papernameS{} uses a four-round Low-Latency Block Cipher (LLBC)\footnote{Any low-latency lightweight block cipher, such as SCARF\cite{SCARF}, can be used.}, similar to CEASER\cite{CEASER} and CUBE\cite{cube}, to encrypt \textit{n}-bit row addresses (e.g., 21-bit for 2M rows). The LLBC takes the original row address as input and uses one key per round, generated at boot time and updated every \TREFW{} (32ms) via a Pseudo-Random Number Generator (PRNG) or True Random Number Generator (TRNG). The keys, stored in four 16-bit registers, ensure unique indices for each row, enhancing security and performance.

\begin{figure}[h]
    \centering
    \includegraphics[width=0.8\linewidth]{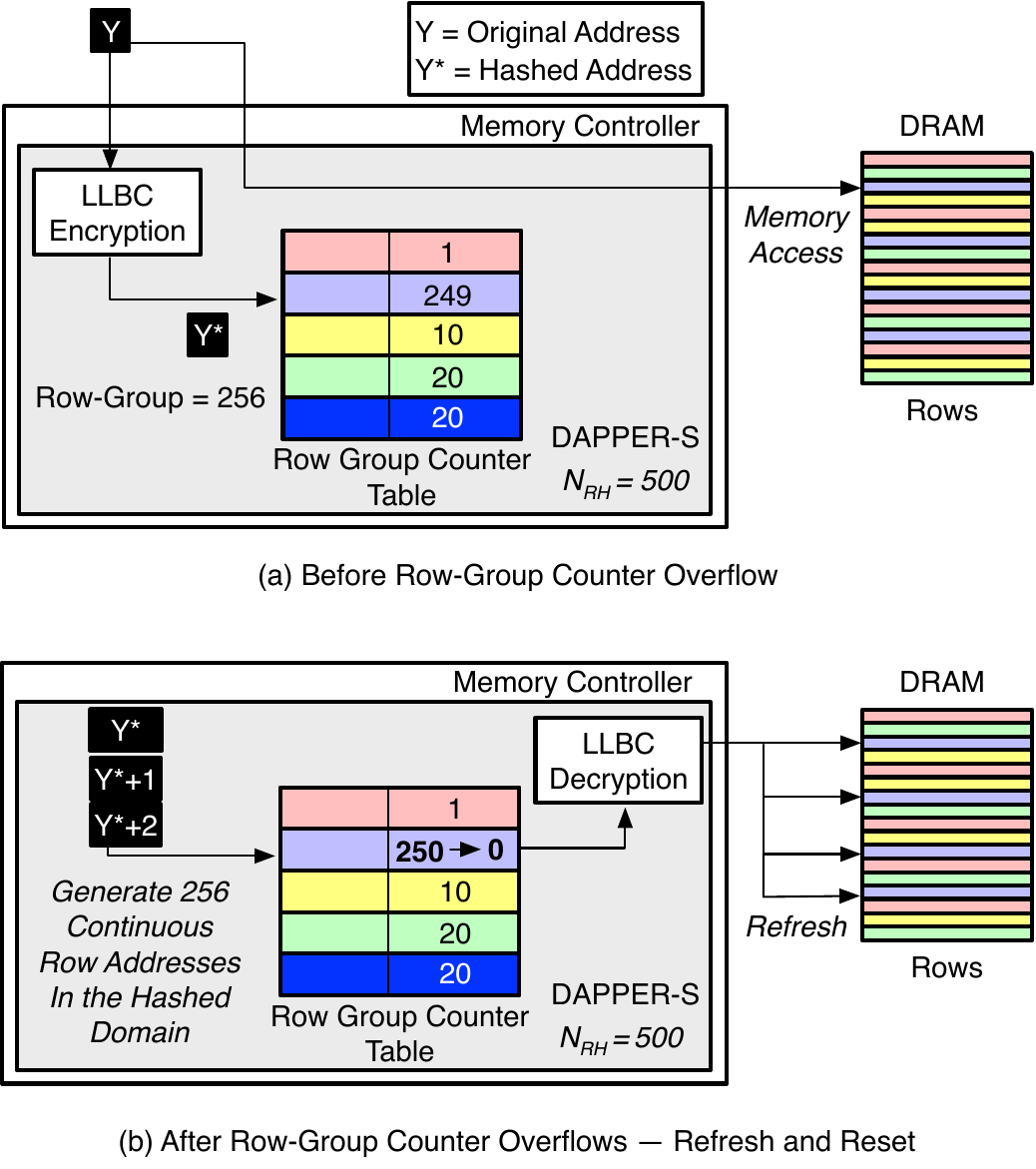}
    \vspace{-0.08in}
    \caption{Overview of \papernameS. The key operation involves updating the Row Group Counter (RGC) during each memory access. (a) Rows are securely hashed to ensure even distribution across RGCs, with each counter tracking 256 rows. (b) When an RGC reaches the mitigation threshold ($\frac{\text{N}_{\text{RH}}}{2}$), \papernameS{} decrypts the hashed addresses to their original addresses, performing mitigative refreshes to all rows in the group, and resets the RGC to zero.}
    \label{fig:DAPPER-S}
    \vspace{-0.1in}
\end{figure}

\subsection{Design of DAPPER-S}
\cref{fig:DAPPER-S} shows the design of \papernameS{}, which incorporates both secure hashing and mitigation mechanisms. To prevent attacks that exploit repetitive DRAM accesses for counter updates (similar to attacks for Hydra and START described in \cref{subsec:Perf-Attacks_on_SOTA}), all RGCs are stored in the memory controller. \NMIT{} is set to half of the RowHammer threshold (\NRH{}) to securely reset RGCs every \TREFW{}, in line with previous studies\cite{scale-srs, hydra,  comet}. The hashing process, shown in \cref{fig:DAPPER-S}(a), randomizes each DRAM row access using the LLBC engine, which transforms the original address (Y) into a hashed address (Y$^{*}$). The hashed address is then divided by the row group size (e.g., 256) to determine the corresponding RGC. The mitigation process is triggered when an RGC reaches \NMIT{}. As depicted in \cref{fig:DAPPER-S}(b), \papernameS{} decrypts the hashed addresses back to their original address, performs mitigative refreshes to all rows in the group, and resets the RGC to zero.

\subsection{Mapping-Capturing Attacks on DAPPER-S}\label{attack-single}
\begin{figure}[h]
    \centering
    \vspace{-0.1in}
    \includegraphics[width=0.6\columnwidth]{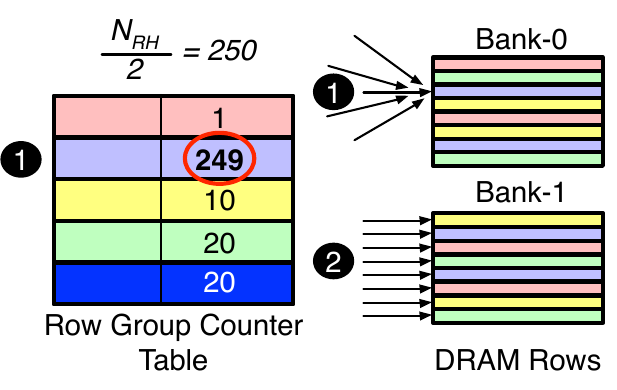}
    \vspace{-0.1in}
    \caption{\MAPPINGATTACK{} attacks, which targets the static hash mapping of \papernameS{}. The attack begins by activating a single row in a bank (e.g., bank-0) to increment the RGC count just below the mitigation threshold (${\text{N}_{M}} - 1$). It then moves to the next bank (e.g., bank-1) and sequentially activates the rows while monitoring the issuance of mitigative refreshes. Upon detection of a refresh, the activated row is identified as belonging to the same RGC. This process is repeated across all banks to uncover the static hash mapping.}
    \label{fig:learning_mapping_DAPPER-S}
    \vspace{-0.05in}
\end{figure}

\cref{fig:learning_mapping_DAPPER-S} illustrates a \MAPPINGATTACK{} attack that targets the static hash mapping of \papernameS{} to identify rows mapped to the same row group and overflow the Row Group Counter (RGC). By inflating RGCs, this attack triggers unnecessary mitigative refreshes with fewer activations than \NMIT{}. It unfolds in two stages: (1) repeatedly activating a single row (or two rows under the open-page policy) just below \NMIT{}{} (N$_{\text{M}} - 1$) (\encircle{1}) and (2) sequentially activating consecutive rows in a different bank until a mitigative refresh occurs\footnote{We assume attackers have exclusive access and can detect timing differences caused by mitigative refreshes.}(\encircle{2}). Capturing even a single pair of mappings enables effective attacks, such as simultaneously activating multiple rows within the same group across different banks to accelerate RGC overflows.

\papernameS{} counters this by resetting the RGC table and updating hash functions during each \tRESET{} period to disrupt attack patterns. For the attack to succeed, attackers must identify at least one mapping pair before \tRESET{} expires. After activating the target row by \NMIT{} - 1 times, the remaining attack time (\tLEFT{}) is calculated using \cref{eq:remaining attack time}.

\vspace{-0.1in}
\begin{equation}\label{eq:remaining attack time}
    \text{t}_{\text{left}} = \text{t}_{\text{reset}} - \textit{tRC} \times (\text{N}_{\text{M}}-1)
\end{equation}

With a 48ns \tRC{} and \NMIT{} of 250, the attack time is approximately 12$\mu$s\footnote{To simplify our analysis, we do not consider the impact of periodic DRAM refresh operations. The required attack time will be longer if we account for the refresh operations.}. Now, the maximum number of activations for each channel (\ACTMAX{}) that can be issued during \tLEFT{} can be calculated by~\cref{eq:act_num}. 

\begin{equation}~\label{eq:act_num}
    \text{ACT}_{\text{MAX}} = \frac{\text{t}_{\text{left}}}{\text{tRRD\_S}}
\end{equation}

The total number of row groups (\NRG{}) is determined by the total number of rows (R) in the randomized space and the row group (RG) size (S), calculated as \NRG{}$=\frac{R}{S}$. With a default RG size of 256 and 2M rows in the randomized space, there are 8K row groups. Each row group can be selected with a probability of $p=\frac{1}{\text{N}{_\text{RG}}}$ for each activation. The attack success probability (P$_\text{S}$) represents the likelihood of selecting the target row group at least once within \ACTMAX{} attempts, as described by \cref{eq:success_probability}.

\vspace{-0.1in}
\begin{equation}\label{eq:success_probability}
    \text{P}_\text{S} = 1 - (1-p)^{\text{ACT}_{\text{MAX}}}
\end{equation}

Based on this, the expected number of attack iterations (\ATITER{}) is represented by~\cref{eq:itteration}.

\vspace{-0.1in}
\begin{equation}\label{eq:itteration}
    \text{AT}_{\text{iter}} = \frac{1}{{\text{P}_\text{S}}}
\end{equation}

Finally, the time (\ATTIME{}) for one successful attack can be calculated using \cref{eq:attacktime}.

\vspace{-0.1in}
\begin{equation}\label{eq:attacktime}
    \text{AT}_{\text{time}} = \text{t}_{\text{reset}} \times \text{AT}_{\text{iter}}
\end{equation}

Table~\ref{tab:attacktime_DAPPER_S} shows the required attack iterations and time to decipher at least one pair of row mappings in \papernameS{}. Even with an impractically short reset period of 12$\mu$s, relying on a single secure hash can be broken every 7.6 ms. This highlights the need for significantly stronger security to protect the mapping information. As a result, we are motivated to enhance \papernameS{} to better defend against \MAPPINGATTACK{} attacks. Our enhanced design targets a 99.99\% prevention rate against \MAPPINGATTACK{} attacks within \TREFW{} (32 ms).

\begin{table}[h]
    \vspace{-0.15in}
    \centering
    \caption{Vulnerability of \papernameS{} to \MAPPINGATTACK{} Attacks}
    \resizebox{3.3in}{!}{
    \begin{tabular}{ccc}\hline
       \textbf{Reset Period ($\text{t}_{\text{reset}}$)}  & \textbf{Attack Iterations (AT$_{\text{iter}}$)} & \textbf{Attack Time (AT$_{\text{time}}$)} \\\hline
       36$\mu$s        &  1.8 &  64$\mu$s \\
       24$\mu$s        &  3 &  71$\mu$s \\
       12$\mu$s        &  630.6 &  7.6ms \\\hline
       
    \end{tabular}
    }
    \label{tab:attacktime_DAPPER_S}
\end{table}

\begin{figure*}[t!]
    \centering
    \includegraphics[width=0.8\linewidth,height=\paperheight, keepaspectratio]{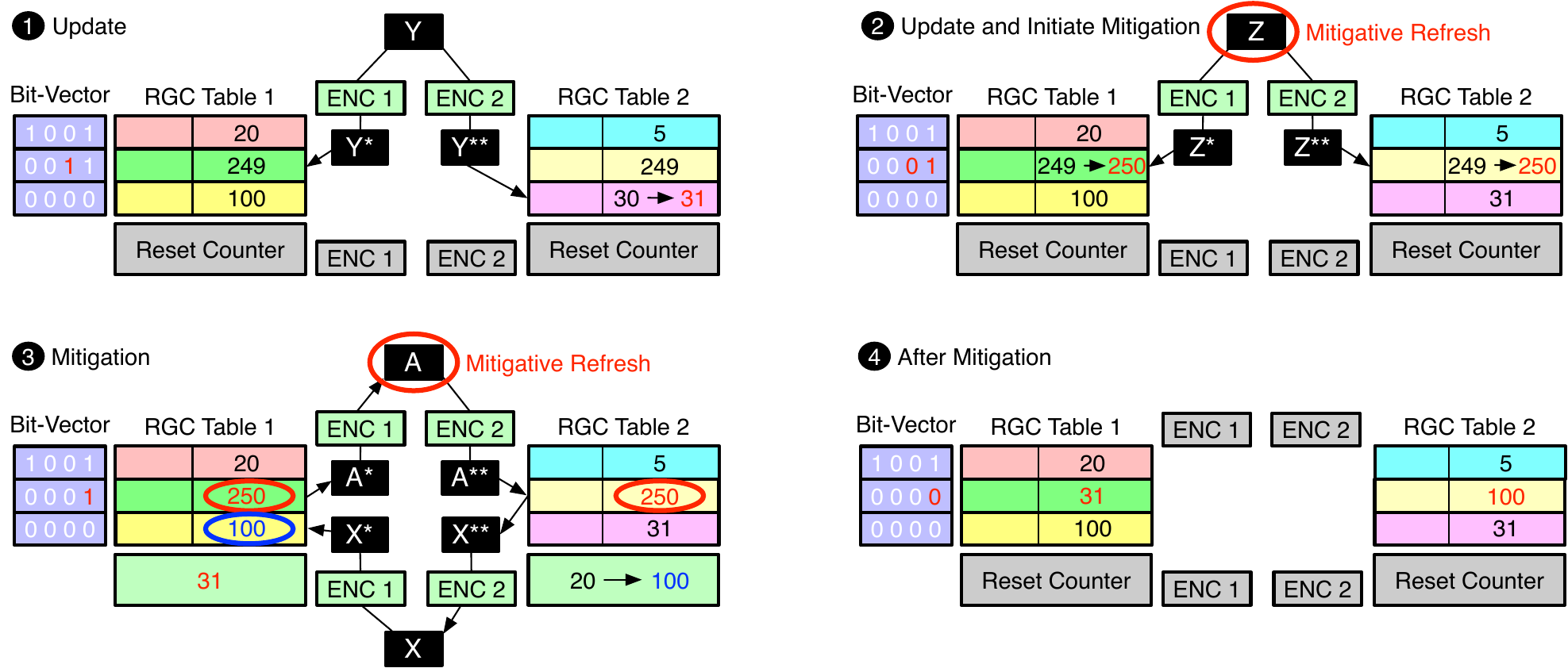}
    \vspace{-0.1in}
    \caption{Overview of \papernameH{}. It uses two hashes to filter out accesses that overwhelm the Row Group Counter (RGC). \papernameH{} performs mitigations only when both RGCs are equal to the Mitigation threshold (\NMIT{}).}
    \vspace{-0.20in}
    \label{fig:DAPPER-h-operations}
\end{figure*}
\begin{figure}[t!]
    \centering    \includegraphics[width=0.45\textwidth,height=\paperheight,keepaspectratio]{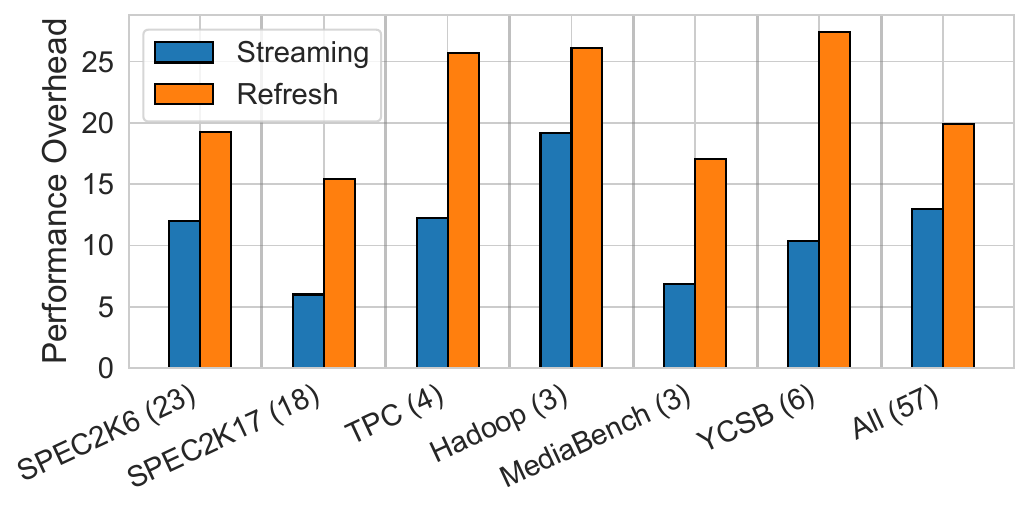}
    \vspace{-0.1in}
    \caption{Performance impact of two  \MAPPINGAGNOGATTACK{} attacks, streaming and refresh attacks, on \papernameS{}. The streaming attack drops performance by 13\%, while the refresh attack leads to a 20\% slowdown.}
    \label{fig:dos_results_dapper_S}
    \vspace{-0.2in}
\end{figure}

\subsection{Mapping-Agnostic Attacks on DAPPER-S}\label{subsec:mapping_agnostic_attack}
While \papernameS{} employs randomized mapping to mitigate \MAPPINGATTACK{} attacks, it remains vulnerable to \MAPPINGAGNOGATTACK{} attacks. One notable example is the \textit{streaming attack}, which sequentially activates all DRAM rows within a rank, frequently triggering mitigative refreshes for each row. For instance, in our baseline system, this attack results in mitigative refreshes across all 2M rows per rank every 6 ms. With a row group size of 256 and \NRH{} of 500, the streaming attack reduces overall performance by 13\%, as illustrated in \cref{fig:dos_results_dapper_S}. Another example is the \emph{refresh attack}, which repeatedly activates single or multiple rows per bank to prompt mitigative refreshes. This attack exploits \papernameS{}'s requirement to refresh all member rows in an RGC (e.g., 256 rows in our default configuration) when its counter reaches \NMIT{}. Consequently, it causes significant slowdowns of 20\%, as shown in \cref{fig:dos_results_dapper_S}. Addressing these \MAPPINGAGNOGATTACK{} attacks is another key goal of our enhanced design.

\section{Enhancing the DAPPER Tracker}\label{sec:design2}
We introduce \papernameH{}, an enhanced tracker designed to defend against both \MAPPINGATTACK{} and \MAPPINGAGNOGATTACK{} attacks while preserving the cost efficiency of \papernameS{}. \papernameH{} employs double-hashing for row group mapping, making it significantly difficult for attackers to capture mapping information. Double-hashing also mitigates the refresh attack, which we will describe in \cref{subsec:DAPPER-H-mapping-agnostic}.

\subsection{Overview of DAPPER-H}
Unlike \papernameS{}, which uses a single Row Group Counter (RGC) table and performs mitigative refreshes to all rows in a group during mitigation, \papernameH{} tracks activations using two distinct RGC tables and performs mitigative refreshes only for the shared rows of the accessed groups. Each table uses a unique hashing function to have different row groupings. Mitigative refreshes are triggered only when the RGCs in both tables reach the Mitigation threshold (\NMIT{}). This approach enhances resilience against Performance Attacks (\PERFATTACKS{}) but introduces additional challenges, such as managing RGC resets during mitigations. We will discuss how these issues are addressed in subsequent sections.

\subsection{Design and Operations of DAPPER-H}
\cref{fig:DAPPER-h-operations} illustrates the overall design of \papernameH{}. It uses two distinct row group counter (RGC) tables, each integrated with a Low-Latency Block Cipher (LLBC) to randomize addresses upon every activation. \papernameH{} incorporates a bit-vector for each entry in RGC Table 1 to further mitigate the streaming attack, where each bit represents a bank. The RGC of Table 1 is incremented only if the corresponding bit in the bit-vector is already set. The following section (\cref{subsec:DAPPER-H-mapping-agnostic}) details how the bit-vector effectively counters the streaming attack. Here, we describe the operations of \papernameH{}. 

\smallskip
\subsubsection{Initialization and Reset}
Like \papernameS{}, \papernameH{} sets \NMIT{} to half of the RowHammer threshold (\NRH{}), ensuring periodic resets of its tracking structures every \TREFW{}. During each reset, both RGC tables and bit-vectors are cleared. These resets occur at the end of each \TREFW{} interval or during system initialization, such as at boot time.

\subsubsection{Update Operation with Bit-Vector}
\papernameH{} uses the per-table LLBC for each activation to compute randomized addresses and accesses the corresponding entries in both RGC tables. For RGC Table 1, \papernameH{} first checks the bit-vector associated with the bank of the activated row. Suppose the bit for the bank is not set; it is updated to 1, and only the RGC in Table 2 is incremented, thereby preventing overestimation of counters due to activations from different banks. For example, if the memory controller activates row Y in Bank 1 and the corresponding bit in the bit-vector is not set, \papernameH{} sets the bit and increments only the RGC in Table 2 (\encircle{1}). 

In contrast, if the bit is already set, both RGCs in Tables 1 and 2 are incremented, and the bit-vector entries for all other banks are cleared. For instance, if the memory controller activates row Z in Bank 0, where the bit for the bank is already set, \papernameH{} increments both RGCs in Tables 1 and 2 and clears the bits for other banks (\encircle{2}). Additionally, since both RGCs reach \NMIT{}, the mitigation process is triggered, and \papernameH{} issues mitigative refreshes to row Z. This entire process, including address randomization and RGC table access, is completed within a single cycle (0.25 ns)\cite{CEASER_S, CEASER, cube}, staying well within the \emph{\tRRDS{}} latency of 2.5 ns.
\begin{figure*}[t!]
    \centering
    \includegraphics[width=0.95\linewidth]{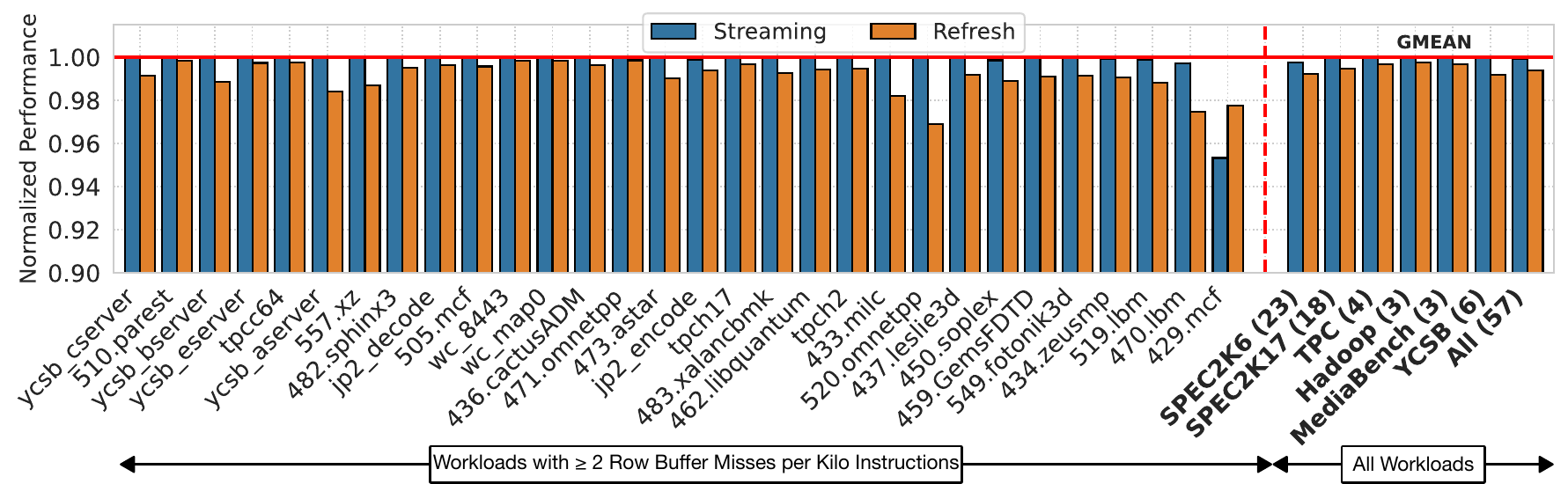}
    \vspace{-0.12in}
    \caption{Normalized performance of \papernameH{} under two \MAPPINGAGNOGATTACK{} attacks, streaming and refresh attacks, at an ultra-low RowHammer threshold (\NRH{}) of 500. The performance of three benign applications is normalized to a non-secure baseline system. Despite the active \PERFATTACKS{}, \papernameH{} incurs less than 1\% average slowdowns. The maximum slowdowns are 4.7\% for the streaming attack and 2.3\% for the refresh attack, highlighting the effectiveness of the proposed double-hashing, novel reset, and bit-vector techniques in mitigating \PERFATTACKS{}.}
    \label{fig:mapping_agnostic_dapper_results}
    \vspace{-0.15in}
\end{figure*}
\begin{figure*}[t!]
    \centering
    \includegraphics[width=0.95\linewidth]{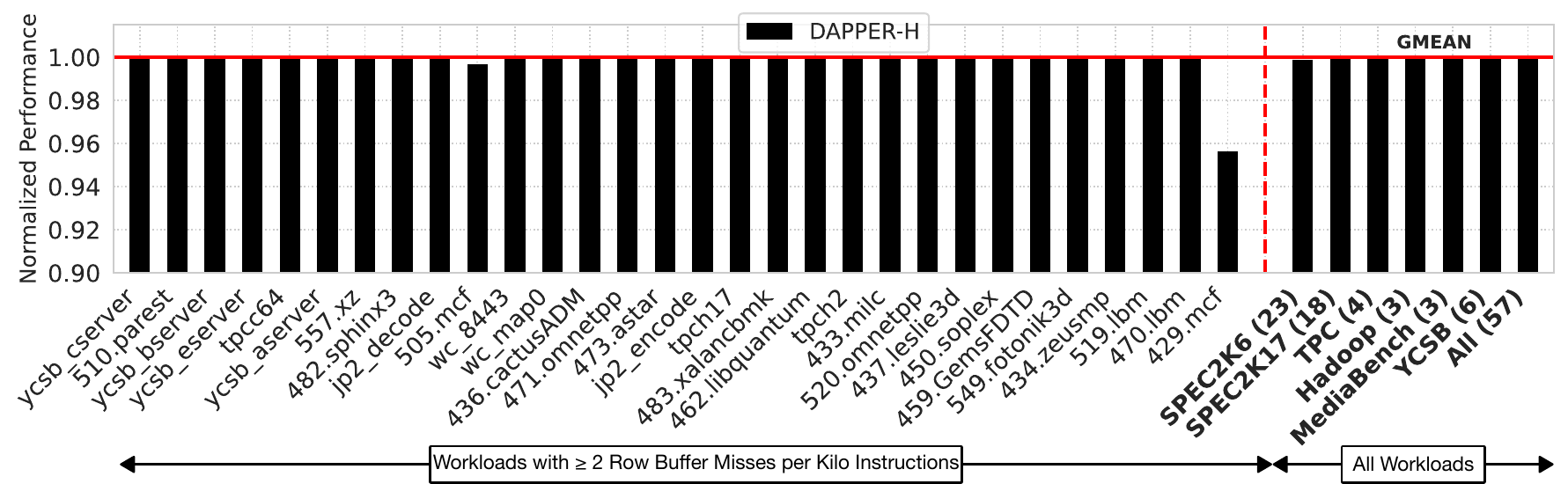}
    \vspace{-0.12in}
    \caption{Normalized performance of \papernameH{} on benign applications compared to an insecure baseline at an ultra-low RowHammer threshold (\NRH{}) of 500. \papernameH{} incurs only 0.1\% slowdown on average and a maximum slowdown of 4.4\%, as the double-hashing, novel reset, and bit-vector mechanisms effectively minimize unnecessary mitigative refreshes.
    }
    \label{fig:benign}
    \vspace{-0.2in}
\end{figure*}

\subsubsection{Mitigative Refreshes and RGC Reset}\label{subsec:dapper_h__mitigation}
Thanks to double-hashing, \papernameH{} avoids issuing mitigative refreshes to all rows in an RGC. Instead, it refreshes only the rows shared between the RGCs of the two tables. This is achieved by decrypting the member rows of each RGC using multiple LLBC engines in parallel and identifying shared rows that require mitigative refreshes (e.g., Row A in \encircle{3}). However, this selective mitigation complicates the counter-reset process, as simply resetting all RGCs to zero is no longer sufficient. To address this, \papernameH{} tracks the counter value of each member row from the opposite RGC table during the decryption process using per-table reset counters. For example, as shown by \encircle{3}, \papernameH{} updates the reset counter value of RGC Table 2 from 20 to 100 because the counter value of member row X in Table 1 is larger than the current reset counter value. At the end of the mitigation process, \papernameH{} resets each RGC to its corresponding reset counter value and clears the bit-vector (as shown by \encircle{4}).

\subsection{Defending Mapping-Capturing Performance Attacks}
Attackers can develop a \MAPPINGATTACK{} attack for \papernameH{} similar to those targeting \papernameS{}. They first choose a single target row and repeatedly activate it. However, due to the double-hashing design of \papernameH{}, they now should activate the target row only \NMIT{} - 2 times. Suppose an attacker were to activate the target row just below the Mitigation threshold (i.e., \NMIT{} - 1 times). In that case, their only chance of learning the mapping relies on randomly selecting a row shared by both row group counters (RGCs). This is highly improbable given the vast number of rows in the randomized address space (e.g., 2M rows in our default configuration). 

A more effective strategy involves activating the target row \NMIT{} - 2 times and randomly accessing two additional rows. The attacker then activates the target row one final time to check if their attack was successful. If the two random accesses correctly target the required RGCs, mitigation will be triggered during the random accesses or the final check activation. This allows the attacker to determine the success of their attack. Given N row groups, the probability of success (P) per trial is described by \cref{eq:sucess}.

\vspace{-0.1in}
\begin{equation}\label{eq:sucess}
    p = (1-(1-\frac{1}{N})^2) \times (1-(1-\frac{1}{N})^2)
\end{equation}

Furthermore, the bit-vector implementation in \papernameH{} prevents attackers from targeting multiple banks with \tRRDS{} latency. Attempts to attack other banks merely set the bit vector and eliminate chances for additional guesses on the target row. This limits attackers to only 616K activations. Additionally, if the attacker fails to capture the mapping, the attacker must activate the target row N$_{\text{M}} - 2$ before retrying. Thus, every attack trial requires the full \NMIT{} limit, reducing the total number of trials (T) to approximately 2.5K. The overall attack success probability is calculated using \cref{eq:succcceess}.

\vspace{-0.1in}
\begin{equation}\label{eq:succcceess}
   P_S = 1 - (1-p)^T 
\end{equation}

As a result, \papernameH effectively prevents \MAPPINGATTACK{} attacks with a 99.99\% probability within \TREFW{}.

\smallskip
\noindent\textbf{Discussion}: Although the mapping information can theoretically be compromised with a 0.01\% probability within \TREFW{} (32ms), attackers would only successfully launch a \MAPPINGATTACK{} attack during \emph{one} \TREFW{} in 10,000 \TREFW{}s (on average). This is because \papernameH{} changes the seed randomly every \TREFW{}. For Performance Attacks (\PERFATTACKS{}) to significantly affect system performance, the attacker must breach a defense \emph{frequently} across back-to-back \TREFW{} intervals. In contrast, RH mitigations require the adversary to achieve a single successful attempt at flipping DRAM bits among millions of tries\cite{rrs}. A single breach can be exploited to manipulate or extract data\cite{seaborn2015exploiting, kwong2020rambleed}, necessitating more complex solutions. Moreover, unlike the case of randomized caches, \papernameH{} does N-to-N encryptions, making it resistant to shortcut attacks\cite{shortcut_llbc1, shortcut_llbc_2} on LLBC. Prior research also demonstrated this aspect\cite{cube}.

\subsection{Defending Mapping-Agnostic Performance Attacks}\label{subsec:DAPPER-H-mapping-agnostic}
\cref{fig:mapping_agnostic_dapper_results} shows the performance of \papernameH{} under two \MAPPINGAGNOGATTACK{} attacks, streaming and refresh attacks, at an ultra-low \NRH{} of 500. The performance of three benign workloads is normalized to an insecure baseline system. \papernameH{} effectively mitigates both attacks, incurring less than 1\% slowdowns on average, with maximum slowdowns of 4.7\% for the streaming attack and 2.3\% for the refresh attack. The double-hashing design of \papernameH{} efficiently counters the refresh attack by performing mitigative refreshes only on shared rows between the two RGCs. Most of the time, it refreshes a single shared row\footnote{The probability of shared rows between two RGCs is very low. With a default 2 million randomized address space and row group size of 256, there are 8K RGCs in each table, making shared rows unlikely. Our evaluation shows that \papernameH{} refreshes a single row 99.9\% of the time under both benign applications and \PERFATTACKS{}, ensuring negligible overhead.}.

As described in \cref{subsec:mapping_agnostic_attack}, the streaming attack that activates every DRAM row can rapidly increase all row group counters (RGCs) in \papernameS{}. This attack exploits bank-level parallelism, activating rows across different banks with \tRRDS{} latency to inflate RGCs. \papernameH{} effectively counters such attack through its bit-vector mechanism, as shown in \cref{fig:mapping_agnostic_dapper_results}. \papernameH{} causes an average slowdown of only 0.2\%, up to 4.7\% in the worst case. Specifically, \papernameH{} sets the corresponding bit in the bit-vector but does not increment the RGC of Table 1 when rows from previously unaccessed banks are activated, preventing the overestimation of RGCs. For example, with a row group size of 256 and uniformly distributed addresses, each bank contains eight rows per group. With our default 2M randomized row space and up to 12 million activations per channel during \TREFW{}, each row is activated approximately six times. Consequently, the theoretical maximum number of accesses per row group is about 48 (6 activations × 8 rows), far below the mitigation threshold (\NMIT{}) of 250. This ensures that \papernameH{} can effectively thwart the streaming attack. While ABACUS~\cite{olgun2023abacus} also uses a per-bank bit-vector for performance reasons, it does not prevent targeted \PERFATTACKS{}. Attackers can still overflow the spillover counter through repeated accesses to rows with different row IDs, leading to significant performance drops.

\subsection{Performance of DAPPER-H on Benign Applications}
\cref{fig:benign} shows the performance of \papernameH{} under benign applications at \NRH{} of 500, normalized to a non-secure baseline DDR5 system. Leveraging double-hashing, novel reset, and bit-vector techniques, \papernameH{} effectively minimizes mitigations, resulting in an average slowdown of only 0.1\%. The highest performance overhead occurs in the most memory-intensive workload, \emph{429.mcf}, which experiences a 4.4\% slowdown due to frequent mitigations than others. Overall, \papernameH{} shows significantly low overheads for benign applications, even at an ultra-low \NRH{} of 500.

\subsection{Sensitivity to Varying RowHammer Thresholds}
\begin{figure}[h!]
    \vspace{-0.1in}
    \centering
    \includegraphics[width=0.45\textwidth,height=\paperheight,keepaspectratio]{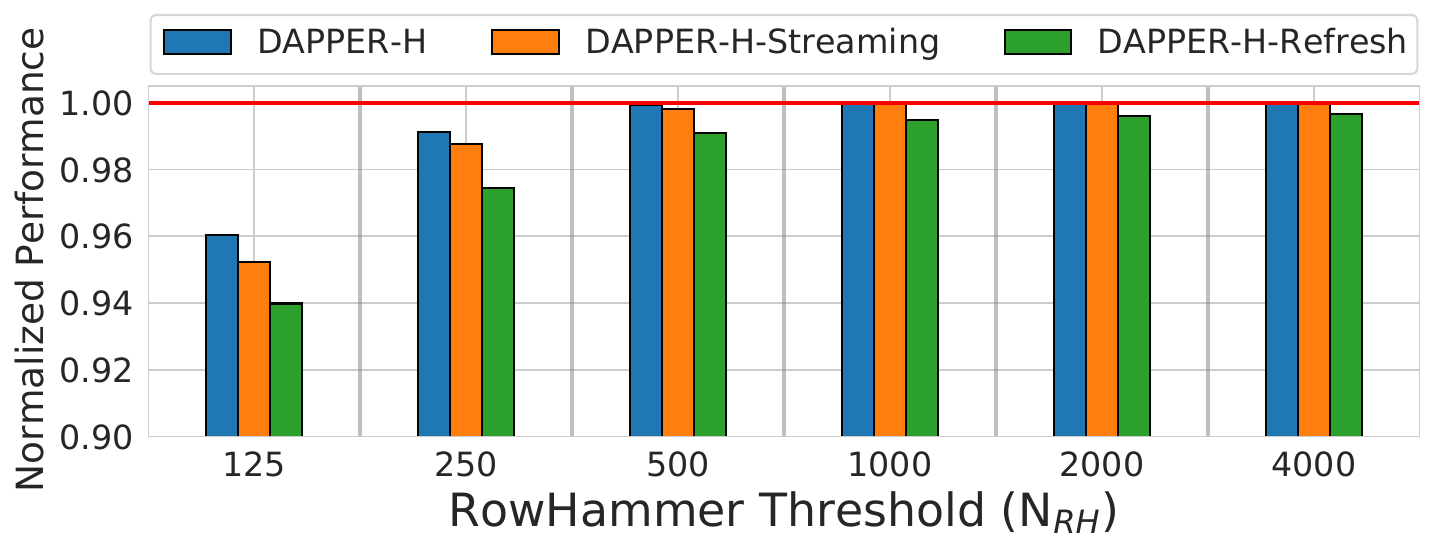}
    \vspace{-0.1in}
    \caption{
    Normalized performance of \papernameH{} under benign and \PERFATTACKS{} as the RowHammer threshold (\NRH{}) varies. \papernameH{} incurs less than 1\% slowdowns at $\text{N}_{\text{RH}} \geq 500$, and up to a 6\% slowdown at \NRH{} of 125 under \PERFATTACKS{}, demonstrating strong resilience against \PERFATTACKS{}.
    }
    \label{fig:rh_sens}
    \vspace{-0.2in}
\end{figure}

\cref{fig:rh_sens} illustrates \papernameH{}'s performance sensitivity as \NRH{} varies from 125 to 4K. \papernameH{} incurs less than 1\% overheads at $\text{N}_{RH} \geq 500$, even under active Performance Attacks (\PERFATTACKS{}). Additionally, it minimizes its performance impact even at lower \NRH{} by avoiding unnecessary mitigative refreshes. For example, at \NRH{} of 125, \papernameH{} incurs only a 6\% slowdown under the refresh attack, highlighting its resilience against \PERFATTACKS{}.

\subsection{Impact of blast radius and Performance with DRFM}\label{subsec:dapper_br_drfm_impact}
\begin{figure}[h!]
    \vspace{-0.1in}
    \centering
    \includegraphics[width=0.45\textwidth,height=\paperheight,keepaspectratio]{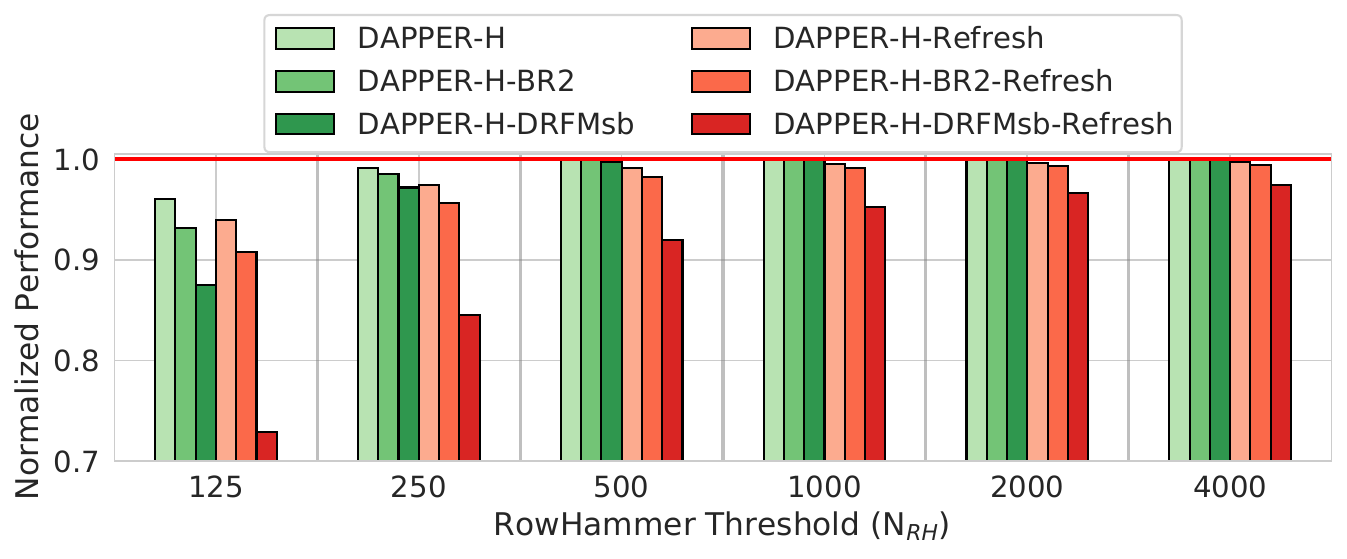}
    \vspace{-0.1in}
    \caption{Normalized performance of \papernameH{} with a blast radius (BR) of 1 (default) and BR of 2, as well as Same-Bank Directed Refresh Management (\DRFMSB{}), under benign applications and the refresh attack. Increasing the BR slightly raises slowdowns while using \DRFMSB{} results in higher slowdowns, particularly at low \NRH{} due to blocking more banks. At \NRH{} of 500 and under the refresh attack, \papernameH{} and \papernameH{}-BR2 incur slowdowns of 1\% and 2\%, respectively, while \papernameH{}-\DRFMSB{} experiences an 8\% slowdown.
    }
    \label{fig:dapper_drfm}
    \vspace{-0.1in}
\end{figure}

\cref{fig:dapper_drfm} compares the performance of \papernameH{} with a blast radius (BR) of 1 (default) and BR of 2 under benign applications and \PERFATTACKS{}. We use the refresh attack for \PERFATTACKS{}, as they cause higher overheads than the streaming attack. Overall, \papernameH{}-BR2 exhibits slightly lower performance. Specifically, it incurs slowdowns of 1.8\% or less at $\text{N}_{\text{RH}} \geq 500$, 4.4\% at \NRH{} of 250, and 9.2\% at \NRH{} of 125 under \PERFATTACKS{}. Conversely, \papernameH{} incurs slowdowns of 0.9\% or less, 2.5\%, and 6\% at the same thresholds. The increased overhead of BR of 2 stems from doubling the blocking duration for each bank during mitigative refreshes.

The JEDEC DDR5 specification supports the Directed Refresh Management (DRFM) command, allowing the memory controller to issue mitigative refreshes for a specific target row, similar to VRR\cite{jedec_ddr5_prac}. However, current DRFM commands support only All-Bank and Same-Bank granularity, blocking either all banks (32 banks) or the same bank across all bank groups (8 banks). This can result in higher overhead than VRR, which blocks only the accessed bank. \cref{fig:dapper_drfm} shows the performance of \papernameH{} with Same-Bank DRFM (\DRFMSB{}). We use \DRFMSB{} with a 240ns delay (supporting a BR of 2) as specified by the JEDEC DDR5 specification\cite{jedec_ddr5_prac}, assuming no rate limitations between DRFM commands\footnote{The JEDEC specification limits the DRFM command rate to one DRFM command per 2 \TREFI{}.}, consistent with prior research\cite{start_hpca24, MINT}. \papernameH{}-\DRFMSB{} incurs higher overhead than \papernameH{}-BR2, as each mitigation penalizes more banks. In the presence of \PERFATTACKS{}, \papernameH{}-\DRFMSB{} causes slowdowns of 8\% at \NRH{} of 500 and 27.1\% at \NRH{} of 125, while \papernameH{}-BR2 incurs only 1.8\% and 4.4\% overheads, respectively, at the same \NRH{}. Our results highlight that supporting per-bank granularity RFM or DRFM commands can significantly reduce the performance overhead, particularly under \PERFATTACKS{}. This observation aligns with findings from concurrent work\cite{qprac}.   

\subsection{Storage and Energy Overhead of DAPPER-H}
\papernameH{} employs double-hashing, which requires dual row group counter (RGC) tables. With the default per-rank mapping, which contains 2M rows in our baseline system, and a Mitigative threshold (\NMIT{}) of 250, \papernameH{} requires 8K 1-byte RGC entries per table, resulting in a total of 32KB SRAM per 32GB DDR5 memory. Additionally, \papernameH{} utilizes a per-bank bit vector to mitigate the streaming attack, which requires 32KB of SRAM per rank or 64KB per 32GB DDR5 memory. Table~\ref{table:areaoverhead} compares \papernameH{}’s storage overhead to state-of-the-art scalable RH mitigations\cite{start_hpca24, hydra, olgun2023abacus, comet} for 32GB DDR5 memory. The estimated die area overhead, calculated based on prior work\cite{olgun2023abacus}, shows that \papernameH{} incurs a negligible 0.038$mm^2$ overhead, similar to existing scalable RH mitigations.

\begin{table}[h!]
    \vspace{-0.1in}
    \centering
    % \vspace{-0.1in}
    \caption{Storage overhead per 32GB DDR5 Memory}
    \resizebox{\columnwidth}{!}{
    \begin{tabular}{cccc} \hline
       & \textbf{SRAM (KB)} & \textbf{CAM (KB)} & \textbf{Die Area Overhead (mm2)} \\\hline
Hydra\cite{hydra}  & 56.5 & -   & 0.044     \\
CoMeT\cite{comet}  & 112  & 23  & 0.139   \\
START\cite{start_hpca24}  & 4    & -   & 0.003     \\
ABACUS\cite{olgun2023abacus} & 19.3 & 7.5 & 0.038     \\
DAPPER-H & 96   & -   & 0.075      \\\hline 
\end{tabular}
}
\label{table:areaoverhead}
\vspace{-0.1in}
\end{table}

\cref{table:energy_overhead} presents the energy overhead of \papernameH{} for \NRH{} values from 125 to 4K, measured using DRAMPower\cite{drampower}. At \NRH{} of 500, \papernameH{} incurs only a 0.1\% overhead for benign applications and a 1.1\% overhead under the refresh attack. Even at \NRH{} of 125, the energy overhead under the refresh attack remains limited to 7.5\%. The overhead primarily arises from mitigation operations (i.e., mitigative refreshes).
\papernameH{} effectively reduces unnecessary mitigations. The double-hashing mechanism ensures the mitigative refreshes target a single shared row 99.9\% of the time, minimizing overhead under the refresh attack. Additionally, the bit-vector mechanism prevents the streaming attack and eliminates redundant mitigations, further optimizing energy efficiency.
\begin{table}[h!]
\centering
\vspace{-0.1in}
\caption{Energy Overhead of \papernameH{}}
\label{table:energy_overhead}
% \vspace{-0.1in}
\resizebox{\columnwidth}{!}{%
\begin{tabular}{cccc}
\toprule
\textbf{RowHammer Threshold} & \textbf{Benign} & \textbf{Streaming Attack} & \textbf{Refresh Attack}\\ \midrule
125             & 4.5\%         &  7.0\%            &  7.5\% \\
250             & 0.9\%         &  1.3\%            &  3.2\% \\
\textbf{500}    & \textbf{0.1\%}&  \textbf{0.2\%}   &  \textbf{1.1\%} \\
1000            & 0\%         &  0.1\%              &  0.6\% \\
2000            & 0\%         &  0\%                &  0.5\% \\
4000            & 0\%         &  0\%                &  0.4\% \\
\bottomrule
\end{tabular}%
}
\vspace{-0.15in}
\end{table}

\begin{figure}[b!]
    \vspace{-0.2in}
    \centering    
    \includegraphics[width=0.45\textwidth,height=\paperheight,keepaspectratio]{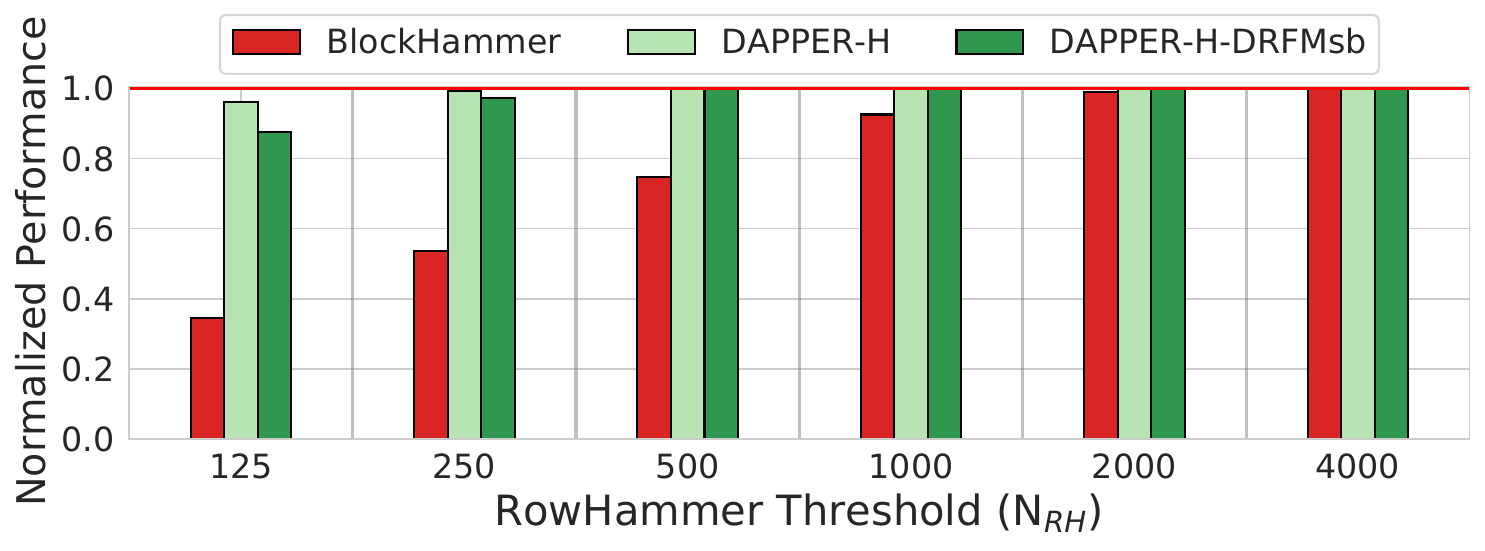}
    \vspace{-0.1in}
    \caption{Performance comparison of \papernameH{} and BlockHammer\cite{yauglikcci2021blockhammer} under benign applications as the RowHammer threshold (\NRH{}) varies. BlockHammer significantly drops performance at ultra-low \NRH{} ($\text{N}_{\text{RH}}\leq500$) due to unnecessary throttling. At \NRH{} of 250, for example, BlockHammer incurs considerable 46.4\% slowdown, while \papernameH{} and \papernameH{}-\DRFMSB{} incurs only 0.9\% and 2.8\% slowdowns, respectively.}
    \label{fig:dappervsblockhammer}
\end{figure}

\subsection{Comparison to BlockHammer}
BlockHammer\cite{yauglikcci2021blockhammer} is a throttling-based RH mitigation that uses a Counting Bloom Filter to track aggressor rows and blacklist them. \cref{fig:dappervsblockhammer} compares the performance of BlockHammer and \papernameH{} under benign applications as \NRH{} varies from 125 to 4K. BlockHammer incurs a 7.5\% slowdown at \NRH{} of 1K and suffers significant overhead at ultra-low \NRH{} (\NRH{} $\leq$ 500). Specifically, it shows slowdowns of 25\% at \NRH{} of 500 and 66\% at \NRH{} of 125. These overheads stem from its inability to accurately distinguish malicious threads, leading to unnecessary throttling of memory requests. This vulnerability can be exploited to launch active \PERFATTACKS{} by hammering rows that share Bloom Filter entries with target rows, effectively blocking the memory requests for target rows and causing severe performance degradation, as highlighted in prior work\cite{mithril, rrs, rubix}. In contrast, \papernameH{} and \papernameH{} with DRFM demonstrate superior scalability. At \NRH{} of 500, they incur less than 0.3\% slowdowns. Even at \NRH{} of 125, \papernameH{} incurs only a 4\% overhead, while \papernameH{}-\DRFMSB{} experiences a 13\% slowdown, significantly outperforming BlockHammer.

\subsection{Comparison to Probabilistic Mitigations}\label{subsec:dapper_vs_prob}
\begin{figure}[h!]
\vspace{-0.1in}
    \centering    
    \includegraphics[width=0.45\textwidth,height=\paperheight,keepaspectratio]{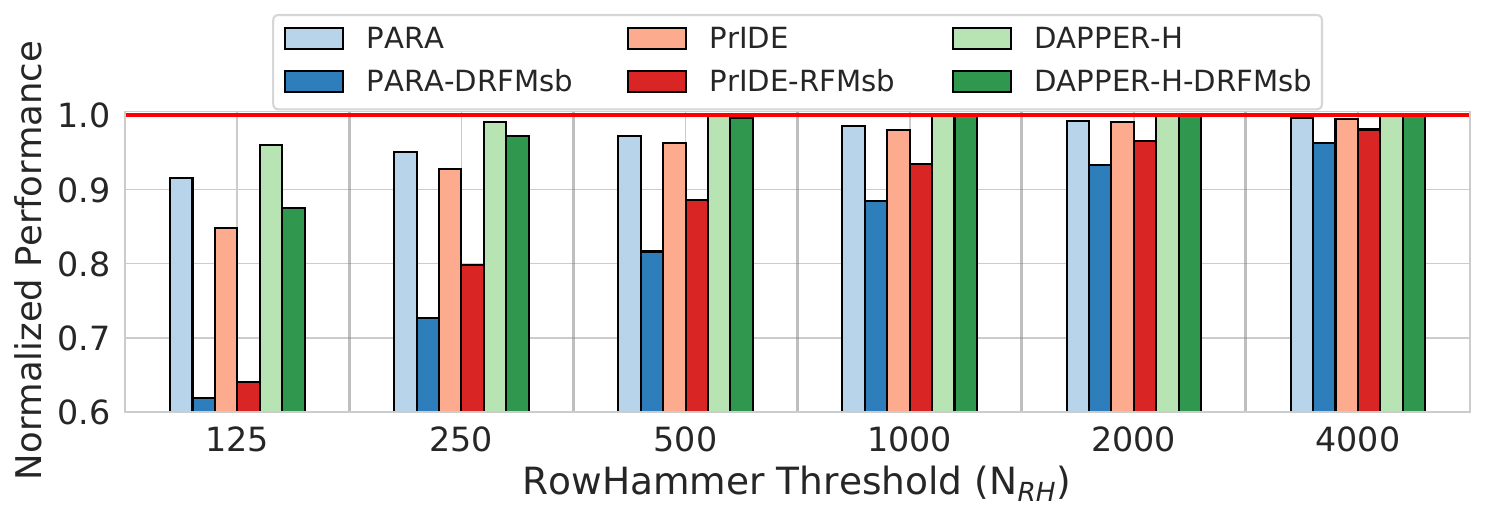}
    \vspace{-0.15in}
    \caption{Normalized performance of \papernameH{}, PARA\cite{kim2014flipping}, and PrIDE\cite{PrIDE} on benign applications as the RowHammer threshold (\NRH{}) varies. At \NRH{} of 500, PARA and PrIDE incur slowdowns of 3\% and 7\%, while PARA-\DRFMSB{} and PrIDE-\RFMSB{} cause significant slowdowns of 18\% and 12\%. Conversely, \papernameH{} and \papernameH{}-\DRFMSB{} incur less than 0.3\% slowdowns.
    }
    \label{fig:dappervsprob_benign}
    \vspace{-0.1in}
\end{figure}

\cref{fig:dappervsprob_benign} compares the performance of \papernameH{} under benign applications to two state-of-the-art probabilistic mitigations, PrIDE\cite{PrIDE} and PARA\cite{kim2014flipping}, as \NRH{} varies from 125 to 4K. All evaluated methods are assumed to support per-bank granularity mitigations by default, with additional comparisons to the existing \DRFMSB{} and \RFMSB{} commands.

Both PARA and PrIDE experience higher slowdowns at ultra-low \NRH{} compared to \papernameH{}. At \NRH{} of 125, PARA and PrIDE incur overheads of 8.5\% and 16.7\%, respectively, whereas \papernameH{} causes only a 4\% slowdown. When paired with Same-Bank RFM (\RFMSB{}) or DRFM (\DRFMSB{}) commands, the slowdowns are even more significant for probabilistic mitigations. At \NRH{} of 500, PARA-\DRFMSB{} and PrIDE-\RFMSB{} incur slowdowns of 18.4\% and 11.5\%, respectively, while \papernameH{}-\DRFMSB{} incurs only 0.3\% overhead. This is due to the stateless nature of probabilistic mitigations, which require more frequent mitigations. In contrast, \papernameH{} minimizes unnecessary mitigations through accurate tracking enabled by double-hashing and bit-vector mechanisms. Additionally, PARA is more affected by Same-Bank mitigations than PrIDE due to the longer delay of the \DRFMSB{} command (240ns) compared to \RFMSB{} (190ns), resulting in greater bandwidth reduction.

\begin{figure}[h!]
    \vspace{-0.1in}
    \centering    
    \includegraphics[width=0.45\textwidth,height=\paperheight,keepaspectratio]{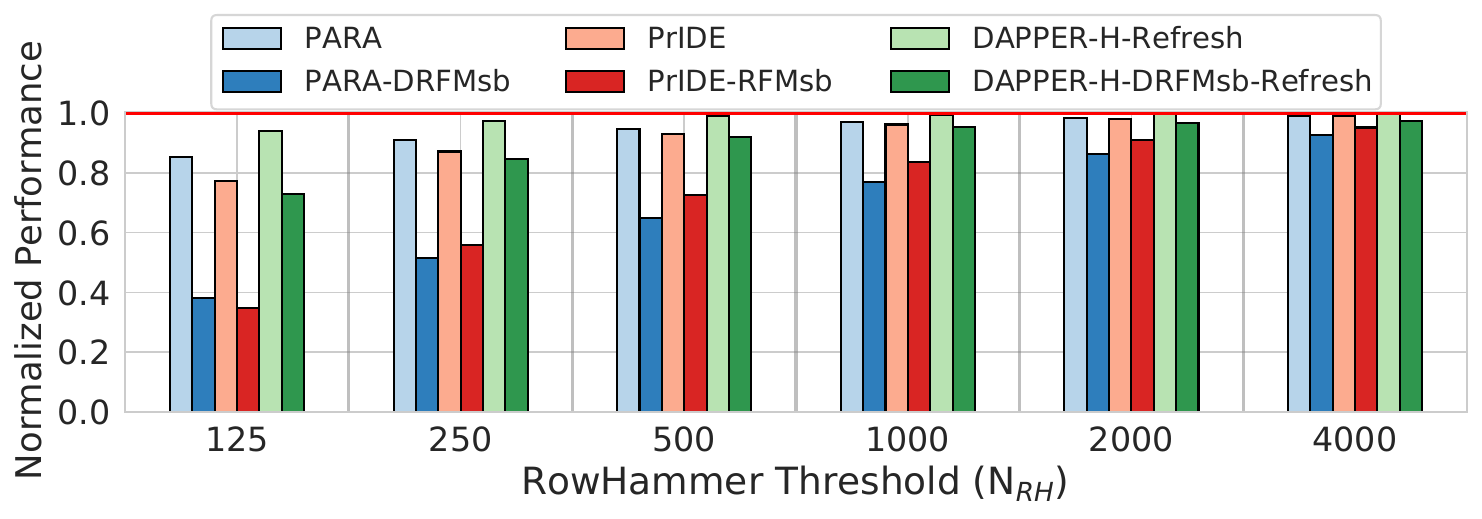}
    \vspace{-0.14in}
    \caption{Normalized performance of \papernameH{}, PARA\cite{kim2014flipping}, and PrIDE\cite{PrIDE} under \PERFATTACKS{}. At the RowHammer threshold (\NRH{}) of 125, \papernameH{} incurs only 6\% overhead, while PARA and PrIDE cause significant slowdowns of 15\% and 23\%, respectively. \papernameH{} minimizes the performance impact of \RFMSB{} and \DRFMSB{} commands more effectively than PARA and PrIDE. At \NRH{} of 1K, \papernameH{}-\DRFMSB{} incurs a 5\% slowdown, while PARA and PrIDE experience considerable slowdowns of 23\% and 16\%, respectively.
    }
    \vspace{-0.18in}
    \label{fig:dappervsprob_perf}
\end{figure}

\cref{fig:dappervsprob_perf} compares the performance of \papernameH{}, PrIDE, and PARA under \PERFATTACKS{}. At \NRH{} of 125, \papernameH{} incurs only a 6\% performance drop, while PARA and PrIDE suffer significant slowdowns of 14.6\% and 22.8\%, respectively, due to frequent mitigations and reduced DRAM bandwidth. These attacks force PARA and PrIDE to perform mitigations every 2 or 4 activations, limiting bandwidth for co-running benign applications. \papernameH{} minimizes the performance impact of Same-Bank mitigations more effectively than probabilistic solutions. At \NRH{} of 1K, \papernameH{}-\DRFMSB{} shows a 4.8\% performance drop, while PARA and PrIDE experience slowdowns of 23\% and 16\%, respectively. These findings underscore the potential benefit of incorporating per-bank RFM or DRFM commands in future DRAM generations to mitigate \PERFATTACKS{}, as discussed in \cref{subsec:dapper_br_drfm_impact}.

\begin{figure}[b!]
    \vspace{-0.2in}
    \centering    
    \includegraphics[width=0.45\textwidth,height=\paperheight,keepaspectratio]{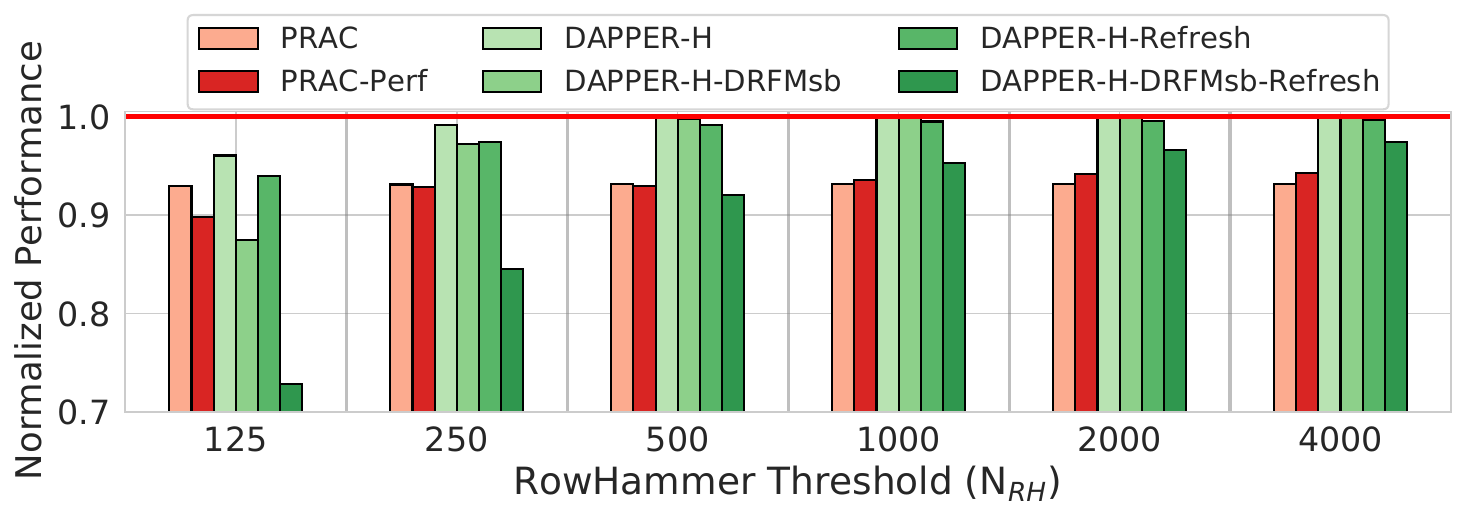}
    \vspace{-0.1in}
    \caption{Normalized performance of \papernameH{} and PRAC under benign applications and \PERFATTACKS{}. While PRAC is less affected by \PERFATTACKS{} due to precise activation counting, it incurs high overhead on benign applications, causing 7\% slowdown even at \NRH{} of 4K. In contrast, \papernameH{} incurs less than 4\% slowdowns for all \NRH{} under benign applications and only 6\% even at \NRH{} of 125 under \PERFATTACKS{}. \papernameH{}-\DRFMSB{} outperforms PRAC at $\text{N}_{\text{RH}} \geq 1\text{K}$ but causes similar or higher overhead at lower \NRH{} due to more frequent mitigations and limited bandwidth due to \DRFMSB{}.
    }
    \label{fig:dappervsprac}
\end{figure}

\subsection{Comparison to Per Row Activation Counting (PRAC)}
The recent JEDEC DDR5 specification\cite{jedec_ddr5_prac} introduces Per Row Activation Counting (PRAC) for precise in-DRAM aggressor tracking, utilizing per-row activation counters and an Alert Back-Off (ABO) protocol. While PRAC can support sub-100 \NRH{}\cite{MOAT,qprac}, it requires read-modify-write operations for every activation to update counters, leading to significant overhead even at relatively high \NRH{} (\NRH{} $\geq$ 1K)\cite{UPRAC}.

\cref{fig:dappervsprac} compares the performance of \papernameH{} and PRAC under benign applications and \PERFATTACKS{}. We implement PRAC based on the state-of-the-art secure QPRAC design\cite{qprac}. PRAC incurs an average of 7\% and up to 20\% overhead on benign applications, even at \NRH{} of 4K, mainly due to frequent counter updates. Its overhead remains relatively constant across evaluated \NRH{} values because per-row tracking minimizes mitigations even at low \NRH{}. Furthermore, PRAC is less impacted by \PERFATTACKS{} at all evaluated \NRH{} values, as its precise aggressor tracking effectively reduces unnecessary mitigative refreshes during \PERFATTACKS{}. These findings align with recent work\cite{UPRAC}.

In contrast, \papernameH{} incurs less than 4\% slowdowns across all evaluated \NRH{} values on benign applications, with only 1\% overhead at $\text{N}_{\text{RH}} \geq 500$ and 6\% at \NRH{} of 125 under \PERFATTACKS{}. While \papernameH{}-\DRFMSB{} outperforms PRAC at $\text{N}_{\text{RH}} \geq 250$ for benign applications and $\text{N}_{\text{RH}} \geq 1\text{K}$ under \PERFATTACKS{}, its overhead increases significantly at $\text{N}_{\text{RH}} \leq 250$ due to more frequent mitigations and greater bandwidth loss caused by \DRFMSB{} commands.

\section{Related Work}
\label{sec:related}
\subsection{RowHammer-Based Performance Attacks}
To the best of our knowledge, our paper is one of the first to thoroughly analyze the vulnerability of existing RowHammer (RH) mitigations to Performance Attacks (\PERFATTACKS{}) and propose a \PERFATTACK{}-resilient solution. BreakHammer \cite{breakhammer2024} is a concurrent work that evaluates the impact of \PERFATTACKS{} on RH mitigations and introduces throttling-based defenses. BreakHammer identifies potentially malicious threads by tracking the number of triggered RH mitigations per hardware thread and throttles their memory requests. However, it depends on prior RH solutions for mitigation and requires microarchitectural modifications, such as changes to caches, to propagate thread information to the throttling logic in the memory controller. In contrast, \papername{} provides standalone \PERFATTACK{} resilience with minimal overhead without requiring additional microarchitectural changes. Furthermore, \papername{} can be combined with BreakHammer to enhance protection against \PERFATTACKS{}.
Another concurrent work, RogueRFM \cite{roguerfm_2025}, also highlights the vulnerability of the current All-Bank Refresh Management (\RFMAB{}) command to \PERFATTACKS{}. RogueRFM proposes limiting the number of \RFMAB{} commands to an average of one per \TREFI{} interval. However, this approach compromises the security of existing RFM-based solutions \cite{PrIDE, MINT, mithril}, which require at least two RFMs to scale to \NRH{} $\leq$ 1.5K.

Several prior studies have acknowledged the concern of RH mitigations against \PERFATTACKS{}. For example, ABACUS\cite{olgun2023abacus} and CoMeT\cite{comet} recognize that their defenses are vulnerable to adversarial patterns similar to those discussed in \cref{subsec:Perf-Attacks_on_SOTA}. However, CoMeT does not offer a solution, and ABACUS suggests increasing the number of Misra-Gries (MG) tracker entries to match the number of rows in the bank. This would require 200KB or 400KB CAM per channel, which is impractical for implementation in the memory controller. In contrast, \papername{} mitigates \PERFATTACKS{} with just 96KB SRAM per 32GB of memory. Additionally, Panopticon \cite{bennett2021panopticon}, UPRAC \cite{UPRAC}, MOAT\cite{MOAT}, and QPRAC\cite{qprac} highlight potential issues with Per Row Activation Counting (PRAC) against \PERFATTACKS{}. MOAT and UPRAC do not provide solutions, while Panopticon proposes randomly initializing per-row activation counters after each mitigation as a solution without any analysis. QPRAC suggests adding Same-Bank or Per-Bank RFM commands for the Alert Back-Off (ABO) protocol based on analytical models. In contrast, our paper provides both through security analysis and evaluation results.

\subsection{Randomization in RowHammer Mitigations}
Recent RH solutions, such as RRS\cite{rrs} or SRS\cite{scale-srs}, mitigate RH by swapping aggressor rows with randomly selected rows\cite{wi2023shadow, scale-srs, rrs} or migrating aggressor rows to a quarantine region\cite{saxena2022aqua}. However, these row randomization techniques remain vulnerable to \PERFATTACKS{}, as adversaries can exploit the high latency of swap or migration operations by triggering frequent mitigations. Rubix\cite{rubix} randomizes physical-to-DRAM address mappings to distribute activations across DRAM rows uniformly. However, it relies on existing RH solutions for mitigation and reduces row-buffer hit rates, potentially incurring high overhead in many workloads. In contrast, \papername{} randomizes \emph{only} counter addresses, not impacting normal DRAM operations such as row-buffer hit rates while providing scalable and standalone security.

\subsection{Host-Side RowHammer Mitigations}
Host-side or memory controller-based RH mitigations address RH either probabilistically \cite{kim2014architectural, kim2014flipping, PROHIT, MRLOC, hammerfilter} or by tracking potential aggressor rows and refreshing their victims\cite{CBT, park2020graphene, hydra, comet, olgun2023abacus, start_hpca24} or throttling access to them\cite{yauglikcci2021blockhammer}. However, at low RH thresholds (\NRH{} $\leq$ 1K), probabilistic solutions incur significant overhead, particularly with the Directed Refresh Management (DRFM) commands. The inefficiency of current DRFM/RFM granularities--blocking access to all banks or the same bank across bank groups--leads to substantial DRAM bandwidth reduction for co-running applications.
Similarly, tracking-based solutions either require impractical storage overhead\cite{park2020graphene, CBT} or remain vulnerable to \PERFATTACKS{}\cite{hydra, comet, olgun2023abacus, start_hpca24}. \papername{} is the only known secure and scalable solution resilient to \PERFATTACKS{} at these ultra-low RH thresholds, even with current DRFM commands.

\subsection{In-DRAM RowHammer Mitigations}
In-DRAM mitigations address RH by refreshing potential aggressor rows during the specific refresh operations (i.e., every N$^{th}$ refresh)\cite{hassan2021uncovering, jattke2022blacksmith, zenhammer_usenex_sec24} or relying on RFMs. These solutions either work probabilistically\cite{dsac_samsung, PAT, PrIDE, MINT} or track potential aggressor rows using counters\cite{lee2019twice, mithril, marazzi2022protrr}. However, they face challenges at ultra-low RH thresholds, including high overhead or impractical storage requirements for tracking structures. Furthermore, as demonstrated in \cref{subsec:dapper_vs_prob}, RFM-based solutions can experience substantial slowdowns under \PERFATTACKS{} with current Same-Bank RFM command (\RFMSB{}). In contrast, \papername{} requires no DRAM modifications and is more scalable with the existing Same-Bank DRFM (\DRFMSB{}) command.

\subsection{Memory Performance Degradation Attacks}
Prior research has proposed both memory Performance Attacks (\PERFATTACKS{})\cite{mutlu2007memoryperfattack,mutlu2007stall} and defenses\cite{mutlu2007stall,subramanian2014blacklisting}. These works address \PERFATTACKS{} through memory request scheduling algorithms, which is orthogonal to our focus on RH-based \PERFATTACKS{}. However, both types of attacks can coexist, potentially exacerbating performance degradation. Thus, addressing both attacks is critical.

\subsection{Randomization in Memory Systems}
Randomization is widely used to enhance the reliability and security of memory systems. For instance, cache randomization techniques\cite{newcache, scattercache, newcache2007,sudoku} like CEASER\cite{CEASER,CEASER_S} and MIRAGE\cite{MIRAGE} randomize line-to-set mappings to defend against conflict-based cache attacks. Similarly, other work\cite{securityRefresh, startGap} leverages randomization in Non-Volatile Memory mappings for wear-leveling.

\section{Conclusion}
RowHammer (RH) vulnerabilities pose a severe threat to modern memory systems. While shared counters and tracking structures offer cost-effective mitigation, they are vulnerable to Performance Attacks (\PERFATTACKS{}) that exploit these structures, reducing DRAM bandwidth and causing performance drops. We address these challenges with secure hashing mechanisms. Our paper introduces \papername{}, a novel low-cost tracker resilient to \PERFATTACKS{}, even at low RH thresholds. We first present \papernameS{}, a tracker template utilizing secure hashing, and further enhance it with \papernameH{}, which integrates double-hashing, advanced reset, and bit-vector strategies. Our security analysis shows that \papernameH{} effectively mitigates \PERFATTACK{} patterns. Experimental results on 57 workloads from six benchmark suites show that \papernameH incurs just a 0.9\% under active \PERFATTACKS{}, requiring only 96KB of SRAM per 32GB of DRAM.

\section{Acknowledgments}
This project is a part of the Systems and Architecture Laboratory (\textit{STAR Lab}) at the University of British Columbia (UBC). We thank the Advanced Research Computing (ARC) Center team at UBC~\cite{sockeye}. We also thank the anonymous reviewers (MICRO 2024 and HPCA 2025) for their invaluable feedback. This work was partially supported by the Natural Sciences and Engineering Research Council of Canada (NSERC) [funding reference number RGPIN-2019-05059] and a gift from Meta Inc. The views and conclusions contained herein are those of the authors. They should not be interpreted as representing the official policies or endorsements of NSERC, the Canadian Government, Meta Inc., or UBC.

%%%%%%% -- PAPER CONTENT ENDS -- %%%%%%%%

%%%%%%%%% -- BIB STYLE AND FILE -- %%%%%%%%
\bibliographystyle{IEEEtranS}
\balance
\bibliography{main}
%%%%%%%%%%%%%%%%%%%%%%%%%%%%%%%%%%%%

\end{document}